\documentclass[a4paper,11pt]{article}
\pdfoutput=1 

\usepackage{jheppub} 

\usepackage[T1]{fontenc} 

\usepackage[utf8]{inputenc}
\usepackage{amssymb}
\usepackage{amsmath}
\usepackage{amsfonts}
\usepackage{graphicx}
\usepackage{color}
\usepackage{xspace}
\usepackage{comment}
\usepackage{hyperref}
\usepackage[normalem]{ulem}
\usepackage[section]{placeins}
\usepackage{afterpage}
\usepackage{slashed}
\usepackage{enumerate}
\usepackage{enumitem}
\usepackage{caption}
\usepackage{subfigure}
\usepackage{float}
\usepackage{ulem}
\usepackage{verbatim}
\usepackage{multirow,rotating}
\usepackage[dvipsnames]{xcolor}

\definecolor{dcolour}{rgb}{.5, .5, .5}
\def\gsim{\raise0.3ex\hbox{$\;>$\kern-0.75em\raise-1.1ex\hbox{$\sim\;$}}}
\def\lsim{\raise0.3ex\hbox{$\;<$\kern-0.75em\raise-1.1ex\hbox{$\sim\;$}}}
\def\gsim{\raise0.3ex\hbox{$\;>$\kern-0.75em\raise-1.1ex\hbox{$\sim\;$}}}
\def\lsim{\raise0.3ex\hbox{$\;<$\kern-0.75em\raise-1.1ex\hbox{$\sim\;$}}}

\newcommand{\ba}[1]{\begin{eqnarray} \label{(#1)}}
\newcommand{\ea}{\end{eqnarray}}

\newcommand{\iab}{\rm ab^{-1}}
\newcommand{\ifb}{\rm fb^{-1}}
\newcommand{\mltp}{{\mkern-2mu\times\mkern-2mu}}

\newcommand{\met}{\slashed{E}_T}

\title{\boldmath Search for the $\gamma Z$ decay mode of heavy photophobic axion-like particles at the LHC}

\author[a]{Zilong Ding}
\author[a]{, Ying-nan Mao}
\author[a,1]{and Kechen Wang\note{Corresponding author.}}

\affiliation[a]{Department of Physics, School of Physics and Mechanics, Wuhan University of Technology,\\430070 Wuhan, Hubei, China}

\emailAdd{zilong.d@whut.edu.cn}
\emailAdd{ynmao@whut.edu.cn}
\emailAdd{kechen.wang@whut.edu.cn}

\abstract{
We assume the coupling of Axion-like particle (ALP) to diphoton $g_{a\gamma\gamma} \sim 0$ and accomplish detailed analyses for the $\gamma Z$ decay mode of such heavy photophobic ALPs at the high luminosity-Large Hadron Collider (HL-LHC). ALPs are produced with two jets via both the $s$-channel vector boson exchange and vector boson fusion processes, with the signal process $pp \to jj\, a (\to \gamma\, Z (\to \ell^+ \ell^-)\, ) $  for $\ell = e, \mu$. Signal and background events are simulated at the detector-level. Preselection criteria target events with one photon, two oppositely charged electrons or muons, and two non-$b$-tagged jets. The ALP mass is reconstructed, and kinematic observables are input into a machine learning-based multivariate analysis for optimal background rejection. Discover sensitivities on the ALP's coupling to di-$W$ boson, $g_{aWW}$, are presented in the mass range from 100 to 4000 GeV at center-of-mass energy $\sqrt{s} = 14$ TeV and integrated luminosity $\mathcal{L} =$ 3 ab$^{-1}$ and 140 fb$^{-1}$. Sensitivities on the production cross section $\sigma (pp \to jj\, a)$ times the branching ratio Br$(a \to \gamma Z)$ are also presented.
}

\begin{document}

\maketitle

\flushbottom

\section{Introduction}
\label{sec:intro}

The search for axions and axion-like particles (ALPs) is a promising avenue in exploring physics beyond the Standard Model (SM), as they may help resolve open questions in particle physics and cosmology. Axions were originally proposed to address the strong CP (Charge-Parity) problem in Quantum Chromodynamics (QCD)~\cite{Dine:2000cj,Kim:2008hd,Kim:2009xp}, where a CP-violating term in the QCD Lagrangian is not reflected in observed neutron electric dipole moments~\cite{Baker:2006ts,Abel:2020pzs}. The Peccei-Quinn (PQ) mechanism~\cite{Peccei:1977hh,Peccei:1977ur} offers a solution, introducing a new global \( \text{U(1)} \) symmetry whose spontaneous breaking produces the QCD axion, effectively canceling the CP-violating term. ALPs extend this idea, representing a broader class of pseudoscalar particles arising from the breaking of a new \( \text{U(1)} \) symmetry at a high energy scale \( \Lambda \). Unlike QCD axions, ALPs are not constrained to solve the strong CP problem, allowing their masses and couplings to vary independently. This flexibility broadens their parameter space, making ALPs appealing for astrophysical and collider-based searches~\cite{Galanti:2022ijh,Choi:2020rgn,Qiu:2024muo}. At colliders, ALPs are typically explored through their interactions with SM gauge bosons (photons, \( Z \)- and \( W \)-bosons), gluons, the Higgs boson, and fermions~\cite{Jaeckel:2015jla, Liu:2017zdh, Steinberg:2021iay, Carra:2021ycg, Bauer:2017ris, Dolan:2017osp, Bauer:2018uxu, Zhang:2021sio, dEnterria:2021ljz, Agrawal:2021dbo, Tian:2022rsi, Ghebretinsaea:2022djg, Antel:2023hkf, Biswas:2023ksj, Lu:2024fxs}. 

Due to the easy detection and low background of final state energetic photons, previous collider experiments have thoroughly explored the ALP's coupling to diphoton, \(g_{a\gamma\gamma}\), for its mass above \(\mathcal{O}(0.1)\) GeV. BESIII studied \(\psi(3686) \to \pi^+ \pi^-\, J/\psi (\to \gamma\, a (\to \gamma\gamma)\,)\) using \(2.71 \times 10^9\) events, setting limits on \(g_{a\gamma\gamma}\) 
for $m_a$
between 0.165 and 2.84 GeV, with the strongest limit \(\sim 0.5\) TeV\(^{-1}\) for \(m_a\) from 0.165 to 1.468 GeV~\cite{BESIII:2022rzz,Jiang:2023lnw}. Belle II analyzed \(e^- e^+ \to \gamma\, a(\to \gamma\gamma)\) at \(\sqrt{s} = 10.58\) GeV and integrated luminosity \(\mathcal{L} = 445\) pb\(^{-1}\), placing constraints on \(g_{a\gamma\gamma}\) for \(m_a\) from 0.2 to 9.7 GeV, with the strongest limit around 1 TeV\(^{-1}\) for \(m_a\) between 0.2 and 5 GeV~\cite{Belle-II:2020jti}. For masses between 10 and 100 GeV, CMS and ATLAS provided limits from light-by-light scattering in lead-lead collisions, with the strongest limit \(\sim 0.1\) TeV\(^{-1}\) from ATLAS at \(\sqrt{s} = 5.02\) TeV and \(\mathcal{L} = 2.2\) nb\(^{-1}\)~\cite{ATLAS:2020hii,CMS:2018erd}. For \(m_a\) between 100 and 2000 GeV, the strongest limits on \(g_{a\gamma\gamma}\) have been derived from diphoton production in proton-proton collisions at the LHC~\cite{dEnterria:2021ljz}. Additionally, for masses below \(\mathcal{O}(0.1)\) GeV, strong constraints are provided by beam dump experiments, astrophysical and cosmic axion measurements, laboratory searches, and other experiments~\cite{Dobrich:2015jyk, NA64:2020qwq, Bauer:2017ris,BESIII:2022rzz,ParticleDataGroup:2024cfk,ALPlimits}.

Current experiments have set strong bounds on \(g_{a\gamma\gamma}\), implying that if ALPs exist, their coupling to diphoton may be suppressed\,\footnote{
The suppression of $g_{a\gamma\gamma}$ can be realized in different theoretical scenarios~\cite{Craig:2018kne, Fonseca:2018xzp, Hook:2016mqo}.
}
, leading to primary interactions with other SM electroweak bosons. Such ALPs, termed ``photophobic ALP'', cannot be produced or decay via the \(a\)-\(\gamma\)-\(\gamma\) vertex, resulting in different and potentially novel collider signals. Besides, the CDF II collaboration recently reported a \(W\)-boson mass measurement~\cite{CDF:2022hxs} inconsistent with the SM prediction and previous experimental values. A recent study~\cite{Aiko:2023trb} suggests that a heavy ALP with \(g_{a\gamma\gamma} \sim 0\) could resolve this discrepancy and improve the global fit for electroweak precision observables, especially for \(m_a > 500\) GeV. Hence, collider searches for heavy photophobic ALPs are also crucial to explore this possibility.

Previous studies on heavy photophobic ALPs at the LHC derive limits by reinterpreting experimental analyses at \( \sqrt{s} = 8 \)~or~13 TeV with restricted luminosities and mass ranges~\cite{Craig:2018kne, Bonilla:2022pxu,Aiko:2024xiv}.
Limits on ALP masses from 40 to 500 GeV were obtained by reinterpreting SM triboson searches in proton-proton collisions at $\sqrt{s}$ = 8 TeV~\cite{Craig:2018kne}, where the ATLAS searches for \( pp \to Z(\to \nu\bar{\nu} / \mu^+\mu^-) \, \gamma\gamma \)~\cite{ATLAS:2016qjc} and \( pp \to W^\pm(\to\mu^\pm \nu) \, W^\pm(\to\mu^\pm \nu) \, W^\mp(\to jj) \)~\cite{ATLAS:2016jeu} were reinterpreted as ALP-mediated signal processes, while the CMS collaboration's limits on SM \( WW\gamma \) and \( WZ\gamma \) production~\cite{CMS:2014cdf} were used to constrain the ALP coupling.
Ref.~\cite{Bonilla:2022pxu} reinterprets CMS analyses on \( pp \to jj VV' \) production at \( \sqrt{s} = 13 \) TeV, where the diboson \( VV' \) includes \( ZZ \)~\cite{CMS:2020fqz}, \( Z\gamma \)~\cite{CMS:2021gme}, \( W^\pm\gamma \)~\cite{CMS:2020ypo}, \( W^\pm Z \)~\cite{CMS:2020gfh}, and \( W^\pm W^\pm \)~\cite{CMS:2020gfh}, with leptonic decays of the \(W\)- and \( Z \)-bosons. The authors change the signal process to be a non-resonant ALP-mediated vector boson scattering process. Their results can be understood as discovery sensitivities on the photophobic ALP with mass below 100 GeV at the HL-LHC with \( \sqrt{s} = 14 \) TeV and \( \mathcal{L} = 3 \) ab\(^{-1}\). 
In Ref.~\cite{Aiko:2024xiv}, LHC analyses at \( \sqrt{s} = 13 \) TeV including  
the SM production \( pp \to W^\pm W^\pm W^\mp \) by CMS with 35.9 fb\(^{-1}\)~\cite{CMS:2019mpq}, 
and the VBF production of a 125 GeV Higgs boson decaying into a photon and a dark photon, \( pp \to jj\, H(\to \gamma \gamma_D) \), by ATLAS with 139 fb\(^{-1}\)~\cite{ATLAS:2021pdg}, 
the gluon–gluon fusion production of a spin-0 and CP-even resonance \( X \), \( g g \to X \to Z(\to \ell^+ \ell^-)\, \gamma \), by ATLAS with 140 fb\(^{-1}\)~\cite{ATLAS:2023wqy} were reinterpreted as the signal processes of 
\( pp \to W^\pm\, a(\to W^\pm W^\mp) \),  
\( jj\, a(\to \gamma\, Z(\to \nu \bar{\nu})) \), 
\( jj\, a(\to \gamma\, Z(\to \ell^+\ell^-)\, ) \), respectively. 
Limits on photophobic ALP are derived up to 3.4 TeV.

Since targets of referred experimental analyses are not heavy photophobic ALPs, 
differences in signal kinematics and background often lead to conservative or potentially invalid results for previous reinterpreting studies.
In this study, we assume \( g_{a\gamma\gamma} \sim 0 \) and accomplish detailed analyses for the \( \gamma Z \) decay mode of a heavy photophobic ALP at the HL-LHC. Key advancements include: (i) incorporating both \( s \)-channel and vector boson fusion production with two jets to capture all relevant signal cross sections; (ii) conducting detector-level simulations to fully capture signal and background kinematics; (iii) using machine learning-based multivariate analysis (MVA) for optimal signal-to-background discrimination; and (iv) presenting results over a wide mass range from 100 to 4000 GeV.

This paper is organized as follows.
Sec.~\ref{sec:theory} lays out the theoretical framework. Sec.~\ref{sec:sig} and Sec.~\ref{sec:SMbg} describe the target signal and potential background processes. The central methodology is presented in Sec.~\ref{sec:analysis}, which covers preselection, reconstruction of the ALP Mass, and MVA for signal-background discrimination. Sec.~\ref{sec:results} shows the main results, followed by a conclusion in Sec~\ref{sec:conclusion}. Important materials and contents are shown in Appendices. 

\section{Theory Models} 
\label{sec:theory}

We consider the ALP couples to the SU(2)$_{\rm L}$ and U(1)$_{\rm Y}$ gauge bosons only. 
The Lagrangian before the electroweak symmetry breaking (EWSB) is shown as~\cite{Georgi:1986df}
\begin{equation}
\mathcal{L}_{\rm ALP} =
\frac{1}{2}\, \partial_{\mu}a\, \partial^{\mu}a
-\frac{1}{2}\, m_{a}^{2}\, a^{2}
- \frac{c_{\widetilde{W}}}{f_{a}}\, a\, W_{\mu\nu}^{b}\, \widetilde{W}^{b, \mu\nu}
- \frac{c_{\widetilde{B}}}{f_{a}}\, a\,B_{\mu\nu}\, \widetilde{B}^{\mu\nu},
\label{eq:L}
\end{equation}
where $X_{\mu\nu}$ with $X = W, B$ represents the field strength tensor for the SU(2)$_{\rm L}$ or U(1)$_{\rm Y}$ groups, and $\widetilde{X}^{\mu\nu} \equiv \frac{1}{2}\, \epsilon^{\mu\nu\alpha\beta} X_{\alpha\beta}$
with the Levi-Civita symbol $\epsilon^{\mu\nu\alpha\beta}$ and $\epsilon^{0123}$ = 1. 
The ALP field and its mass are represented by $a$ and $m_a$, and $f_{a}$ is the ALP's decay constant. 
There are three independent coefficients, $m_{a}$, 
$c_{\widetilde{W}}/f_a$ and $c_{\widetilde{B}}/f_a$. 
After the EWSB, the interactions are generally expressed as
\begin{equation}
\mathcal{L}_{\mathrm int} = -\frac{1}{4}\, g_{a\gamma\gamma}\, a F_{\mu\nu} \widetilde{F}^{\mu\nu} -\frac{1}{2}\, g_{a\gamma Z}\, a Z_{\mu\nu} \widetilde{F}^{\mu\nu}
-\frac{1}{4}\, g_{aZZ}\, a Z_{\mu\nu} \widetilde{Z}^{\mu\nu} - \frac{1}{2}\, g_{_{aW W}}\, a W_{\mu\nu}^{+} \widetilde{W}^{-\mu\nu}+\cdots.
\label{eq:L1}
\end{equation}

The coupling constants are related to $c_{\widetilde{W}}$, $c_{\widetilde{B}}$, and $f_a$, and they can be expressed as
\begin{eqnarray}
g_{a\gamma\gamma} &=& \frac{4}{f_{a}}\, (s^{2}_{\theta}\, c_{\widetilde{W}} + c^{2}_{\theta}\, c_{\widetilde{B}}), \label{eqn:gaGG} \\
g_{aZ Z} &=& \frac{4}{f_{a}}\, (c^{2}_{\theta}\, c_{\widetilde{W}} + s^{2}_{\theta}\, c_{\widetilde{B}}),\\
g_{a\gamma Z} &=& \frac{2}{f_{a}}\, s_{2\theta}\, (c_{\widetilde{W}}-c_{\widetilde{B}}),\\
g_{_{aWW}} &=& \frac{4}{f_{a}}\, c_{\widetilde{W}}.
\end{eqnarray}
Here, $c_{\theta} = \cos\theta$, $s_{\theta} = \sin\theta$, and $s_{2\theta} = \sin2\theta$  with the
Weinberg angle $\theta$. 
As mentioned above, in this work, we consider the photophobic ALP which does not couple to di-photon at the tree level, i.e. $g_{a\gamma\gamma}\propto s^{2}_{\theta}c_{\widetilde{W}}+c^{2}_{\theta}c_{\widetilde{B}}$ = 0.  
Therefore, 
$c_{\widetilde{B}}$ and $c_{\widetilde{W}}$ are not independent, 
as well as $g_{aZ Z}$, $g_{a\gamma Z}$ and $g_{_{aWW}}$.
With $t_\theta = \tan \theta$, their relations are expressed as 
\begin{eqnarray}
c_{\widetilde{B}} &=& -\, t_\theta^2\, c_{\widetilde{W}}, \\
g_{aZZ} &=& (1 - t_\theta^2)\, g_{_{aWW}}, \\
g_{a\gamma Z} &=& t_\theta\, g_{_{aWW}}.
\end{eqnarray}

\section{Signal production}
\label{sec:sig}

\begin{figure}[h]
\centering
\includegraphics[width=12cm,height=5cm]{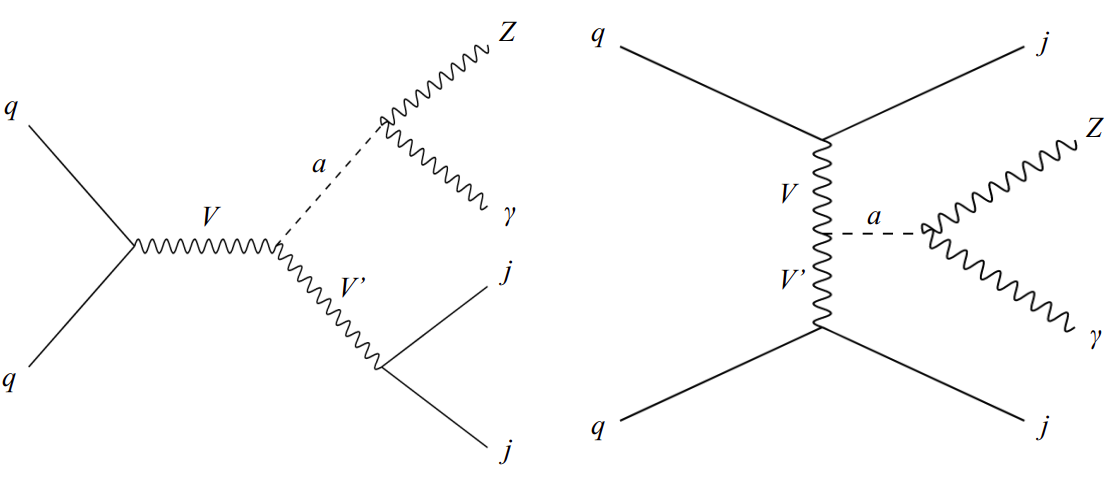}  
\caption{
The $s$-channel (left) and VBF (right) productions of heavy photophobic ALPs associated with two jets at $pp$ colliders, where $a$-$V$-$V'$ vertices include the $a$-$W$-$W$, $a$-$Z$-$Z$ and $a$-$\gamma$-$Z$.
The ALP then decays into a photon plus a $Z$-boson, leading to the signal process of $pp \to j j\, a(\to \gamma Z)$.
}
\label{fig:signal}
\end{figure}

Although the photophobic ALP does not couple to di-photon at the tree level, their interactions to other di-boson (including $a$-$W$-$W$, $a$-$Z$-$Z$ and $a$-$\gamma$-$Z$) still exist.
Such heavy ALPs can be produced via these vertices.
Fig.~\ref{fig:signal} shows the signal processes for the heavy photophobic ALPs considered in this study, where 
ALPs are produced associating with two jets at the HL-LHC via the $s$-channel exchange of vector bosons or vector boson fusions.
The ALP then decays into a photon plus a $Z$-boson.
To suppress the multi-jet background at the $pp$ collider, 
the final state $Z$ bosons is required to decay into a pair of leptons.
The final state contains one photon, two oppositely charged leptons, and at least two jets.
For the photophobic ALP, this signal process has large production cross section, and relatively small background 
at the $pp$ collider.

To generate the signal events, 
we implement the ALP model file with the linear Lagrangian~\cite{Brivio:2017ije} in the Universal FeynRules Output (UFO) format~\cite{Degrande:2011ua} into the MadGraph5\_aMC\,$@$\,NLO program with version 2.6.7~\cite{Alwall:2014hca} to simulate the $pp$ collision, 
where the default ``nn23lo1'' parton distribution function of the proton is used. 
The PYTHIA program with version 8.3~\cite{Bierlich:2022pfr} is utilized to perform the parton showering and the hadronization, and handle the decays of unstable SM particles.  
The CMS
configuration card file is implemented to the Delphes program version 3.4.2~\cite{deFavereau:2013fsa} to complete the detector simulation.
For the jet reconstruction, jets are classified using the FastJet package~\cite{Cacciari:2011ma} with anti-$k_t$ algorithm and cone size $R$ = 0.4.

\begin{figure}[h]
\centering
\includegraphics[width=12cm,height=8cm]{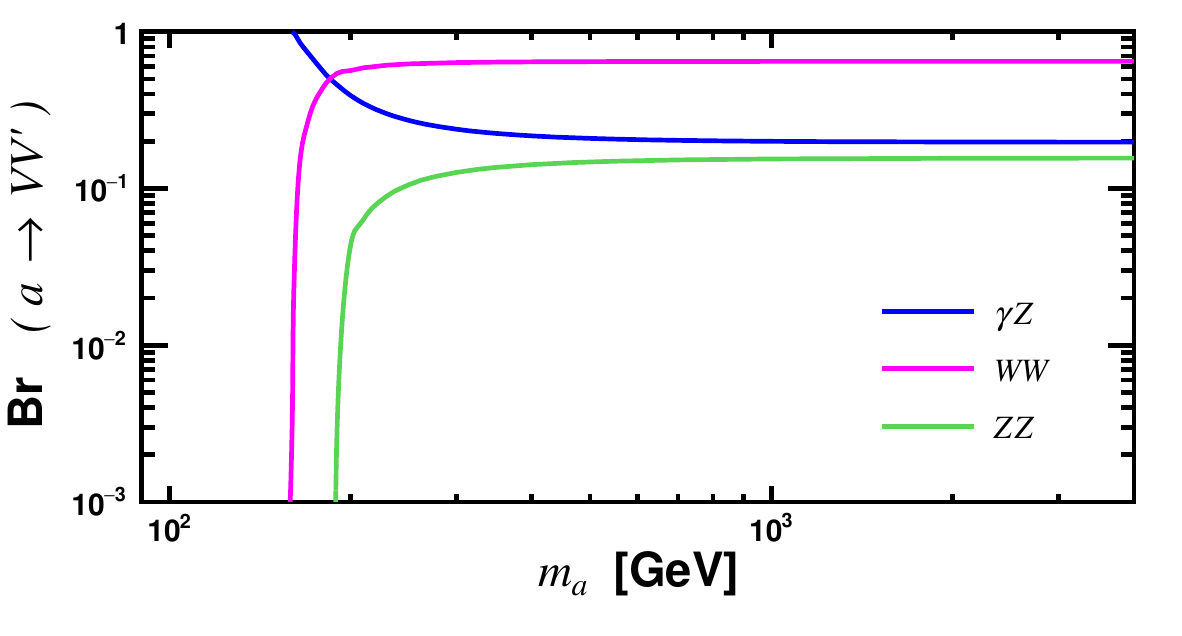}  
\caption{
Branching ratios of dominant decay modes for the photophobic ALP.
}
\label{fig:ratio}
\end{figure}

We calculate the branching ratios and production cross sections using the MadGraph5 program. 
When its mass is above the threshold, the photophobic ALP can mainly decay to di-boson including the modes of $\gamma Z, W^+W^-$, and $ZZ$.
Fig.~\ref{fig:ratio} shows the numerical results of branching ratios, which are checked and founded to be consistent with those in Ref.~\cite{Aiko:2023trb}. 

We note that, in principle, despite the tree level coupling $g_{a\gamma\gamma}$ = 0, the photophobic ALP can still generically decay to di-photon via the loop effects~\cite{Bauer:2017ris,Bauer:2020jbp,Aiko:2023trb}.
Besides, after considering the loop and renormalization group (RG) running effects, 
non-zero branching ratios to fermions (quarks and leptons) and gluons can also be produced.
In Appendix \ref{app:RGrunning}, we calculate the branching ratios of these rare decay channels, and find their numerical results are less than $10^{-2}$ in the considered $m_a$ region.
Therefore, modifications on the tree-level branching ratios due to such effects are tiny and we just consider the tree-level branching ratios in this study.

\begin{figure}[h]
\centering
\includegraphics[width=12cm,height=8cm]{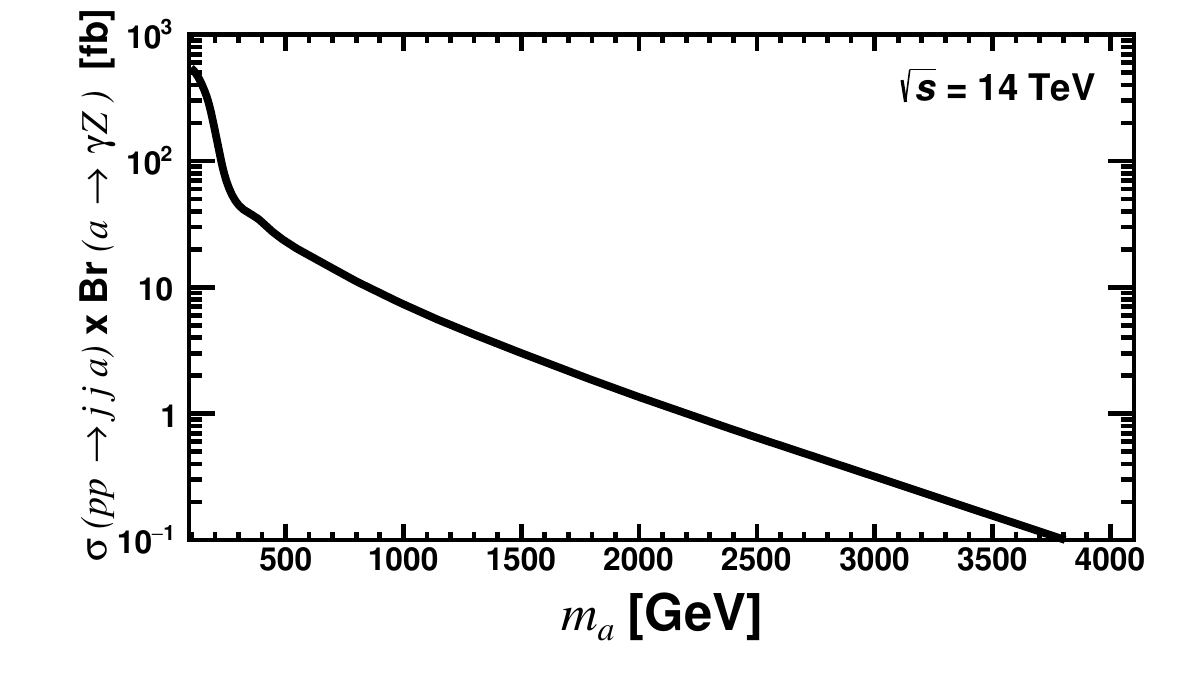}
\caption{
The production cross section of the signal $pp \to j j\, a (\to \gamma Z)$ as varying the ALP's mass at the HL-LHC with $\sqrt{s} =$ 14 TeV, when couplings $g_{a\gamma\gamma}= 0$ and  $g_{_{aWW}}$ = 1 TeV$^{-1}$. 
}
\label{fig:cro}
\end{figure}

In Fig.~\ref{fig:cro}, we plot the cross sections of signal process $pp \to j j a (\to\gamma Z)$ as a function of the ALP's mass at the HL-LHC with center-of-mass energy  $\sqrt{s} =$ 14 TeV, where the $a$-$\gamma$-$\gamma$ coupling $g_{a\gamma\gamma}$ is set to 0, and $a$-$W$-$W$ coupling $g_{_{aWW}}$ is set to 1 TeV$^{-1}$.
The production cross section of this signal process gradually decreases with the increase of the ALP's mass.
The sharp decrease around $m_a \sim 200$ GeV is mainly due to the ALP’s branching ratio to photons and Z bosons dropping rapidly from 160 GeV to 250 GeV (c.f. Fig.~\ref{fig:ratio}).
The contribution to the total cross section from the VBF process increases as the ALP mass increases.
The signal events have two important features: 
(i) the resonance in the reconstructed ALP's mass from the system of photon and dilepton; 
(ii) a pair of extraordinarily forward and backward jets for the VBF process. 

When generating the signal events, we specify that the final state $Z$-boson decays to leptons including flavors of electron, muon and tau.
Thus, the complete signal process is $pp \to j j\, a (\to \gamma\, Z(\to \ell^{+} \ell^{-}) )$ with $\ell = e, \mu$.
Since the kinematics changes as the ALP mass $m_a$, we generate data samples for representative $m_a$ with following values: 100, 165, 250, 400, 700, 1000, 1500, 2000, 2500, 3000, 3500, 4000 GeV.
For each ALP mass, the coupling $g_{_{aWW}}$ is set to be benchmark value of 1 TeV$^{-1}$, 
and at least $10^6$ events are generated.

\section{Background processes}
\label{sec:SMbg}

Since the final state contains one photon, a pair of oppositely charged leptons plus two jets, 
there are mainly six relevant SM background processes: $Z W \gamma$, $Z(\to l^+l^-)\, \gamma j j$, $Z Z \gamma$, $W^+ W^- \gamma$, $W(\to l\nu)\, \gamma jj$, $t \bar{t} \gamma$, where $l = e, \mu, \tau$ and $\nu$ is the SM neutrino or antineutrino. 
We label them as ``B1-B6'' respectively in this article.
When a $Z$-boson decays to a pair of oppositely charged leptons and the remained heavy $W/Z$-boson decays to di-jet, B1-B3 can have the same final state as the signal, and thus, act as the irreducible background.
B4-B5 can contribute as existence of a misidentified lepton.
Moreover, when the final state two $W$-bosons both decay into leptons, B6 can also contribute.

Background events are simulated by importing the SM using the same production programs and procedures as the signal, as described above.
To maintain consistency through this study, the production cross sections calculated by the MadGraph5 program are used to estimate the number of events for both the signal and background processes.
Production cross sections of background processes can be derived from the third column of Table~\ref{tab:presel}. 
To reduce the statistical uncertainty, it is helpful to generate as many background events as possible, 
and we generate data samples within the allowance of our computational resources.
In this study, we produce 13.0 million $Z W \gamma$, 52.0 million $Z(\to l^+l^-)\, \gamma j j$, 11.0  million $Z Z \gamma$,  8.3  million $W^+ W^- \gamma$, 48.8  million $W(\to l\nu)\, \gamma jj$, and 21.1  million  $t \bar{t} \gamma$ events, respectively. Because of their huge production cross sections, B2 and B5 are the main sources of background. 
In order to generate more effective background data, when producing B2 and B5 events, as we have emphasized using the symbols, we specify that the final state $Z/W$-boson decays to leptons, i.e. $Z(\to l^+ l^-)$ and $W(\to l \nu)$ with $l = e, \mu,\tau$. 
 
\section{Data analyses} 
\label{sec:analysis} 

\subsection{Preselection}
\label{subsec:presel}

The final state objects (photons, leptons and jets) are sorted by their transverse momenta and labeled as $O_i,~ i=1,2, ...$ with $O = \gamma,\, \ell,\, j$, respectively.
We apply following preselection criteria
to select the specified final state and reject background events at the first stage.

(i) Events are required to have at least one photon, i.e. $N(\gamma) \geq 1$, and the transverse momentum ($p_{_{\rm T}}$) of the first leading photon is required to be greater than 10 GeV, i.e. $p_{_{\rm T}} (\gamma_{_1}) > 10$ GeV.

(ii) Exactly two leptons (including electrons and muons) with opposite charge and same flavor, i.e. $N(\ell) = 2$ and selecting 
$\ell^+ \ell^-$ with $\ell = e / \mu$; 
$p_{_{\rm T}}$ of first two leading leptons are required to be greater than 10 GeV, i.e. $p_{_{\rm T}}(\ell_1)$, $p_{_{\rm T}}(\ell_2) >$ 10 GeV;

(iii) At least two jets, i.e. $N(j) \geq 2$, and 
$p_{_{\rm T}}$ of the first two leading jets are required to be greater than 30 GeV, i.e. $p_{_{\rm T}}(j_1), p_{_{\rm T}}(j_2) > 30$ GeV.

(iv) The number of $b-$ and $\tau-$tagged jets are required to be zero, i.e. $N(j_b) = 0$ and $N(j_\tau) = 0$.

\begin{table*}[h]
\centering 
\scalebox{0.9}{
\begin{tabular}{ccccccc}
\hline
\hline
$m_a$ [GeV] & 100 & 165 & 250 & 400 & 700 & 1000 \\
&$5.96\mltp10^{-2}$ &$1.30\mltp10^{-1}$ &$1.77\mltp10^{-1}$ &$2.15\mltp10^{-1}$ &$2.49\mltp10^{-1}$&$2.65\mltp10^{-1}$ \\
\hline
$m_a$ [GeV] & 1500 & 2000 & 2500 & 3000 & 3500 & 4000 \\
&$2.81\mltp10^{-1}$ &$2.88\mltp10^{-1}$&$2.90\mltp10^{-1}$  &$2.89\mltp10^{-1}$ &$2.82\mltp10^{-1}$&$2.76\mltp10^{-1}$\\
\hline
background & B1 & B2 & B3 & B4 & B5 & B6 \\
& $ Z W \gamma$ & $ Z(\to l^+l^-) \gamma j j$ & $ Z Z \gamma$& $W^+ W^- \gamma$ & $W(\to l\nu) \gamma jj$ & $ t\bar{t} \gamma $ \\
 & $1.88\mltp10^{-3}$ & $6.91\mltp10^{-3}$ & $3.39\mltp10^{-3}$ & $3.24\mltp10^{-4}$ & $7.6\mltp10^{-6}$ & $1.08\mltp10^{-3}$  \\
 \hline
\hline
\end{tabular}
}
\caption{
Preselection efficiencies for both signals with representative ALP masses and background processes at the HL-LHC with $\sqrt{s} =$ 14 TeV. 
}
\label{tab:preselection}
\end{table*}

Table~\ref{tab:preselection} shows selection efficiencies after applying all preselection criteria (i)-(iv) for both signals with representative ALP masses and background processes at the HL-LHC with center-of-mass energy $\sqrt{s} =$ 14 TeV.
One observes that, the preselection rejects B5 very effectively, with the selection preselection efficiency around $7.6 \mltp 10^{-6}$.
Due to the low misidentified rate of lepton, preselection efficiency of B4 is also small to $3.24 \mltp 10^{-4}$.
Preselection efficiencies of other background processes are slightly larger, $\sim 10^{-3}$. 

\begin{table}[h]
\centering
\begin{tabular}{ccccccc}
\hline
\hline
\multicolumn{2}{c}{HL-LHC} & initial & (i) & (ii) & (iii) & (iv)  \\
\hline
Signal & $jj a(\to \gamma Z(\to l^+ l^-))$ & $4.16\mltp10^{3}$ & $3.56 \mltp 10^{3}$ & $1.35 \mltp 10^{3}$ & $1.17 \mltp 10^{3}$  & $1.03 \mltp 10^{3}$ \\ 
\multicolumn{2}{c}{Total background} & $3.04\mltp10^{9}$ & $4.12 \mltp 10^{8}$ & $8.33 \mltp 10^{6}$ & $2.62 \mltp 10^{6}$  & $2.29 \mltp 10^{6}$ \\ 
\hline
B1 & $ Z W \gamma$ & $1.25\mltp10^{6}$ & $2.47 \mltp 10^{5}$ & $6.51 \mltp 10^{3}$ & $2.67 \mltp 10^{3}$  & $2.35 \mltp 10^{3}$ \\  
B2 & $ Z(\to l^+ l^-) \gamma j j$ & $3.21\mltp10^{8}$ & $4.97 \mltp 10^{7}$ & $8.14 \mltp 10^{6}$ & $2.50 \mltp 10^{6}$  & $2.21 \mltp 10^{6}$ \\ 
B3 & $ Z Z \gamma$ & $5.17\mltp10^{5}$ & $1.07 \mltp 10^{5}$ & $5.17 \mltp 10^{3}$ & $2.24 \mltp 10^{3}$  & $1.71 \mltp 10^{3}$ \\ 
B4 & $W^+ W^- \gamma$ & $3.70\mltp10^{6}$ & $7.39 \mltp 10^{5}$ & $1.60 \mltp 10^{4}$ & $1.26 \mltp 10^{3}$  & $1.14 \mltp 10^{3}$ \\ 
B5 & $W(\to l\nu) \gamma jj$ & $2.97\mltp10^{9}$ & $3.93 \mltp 10^{8}$ & $6.12 \mltp 10^{4}$ & $2.55\mltp 10^{4}$  & $2.25 \mltp 10^{4}$ \\ 
B6 & $ t\bar{t} \gamma$ & $1.72\mltp10^{7}$ & $4.67 \mltp 10^{6}$ & $1.03\mltp 10^{5}$ & $8.75 \mltp 10^{4}$  & $1.35 \mltp 10^{4}$ \\ 
\hline
\hline
\end{tabular}
\caption{
Number of events for the signal with benchmark $m_a$ = 700 GeV and background processes, after applying preselection criteria (i)-(iv) sequentially. 
Here, the numbers correspond to the HL-LHC with $\sqrt{s} =$ 14 TeV and $\mathcal{L} = 3\,\, \iab$.
}
\label{tab:presel}
\end{table}

To show effect of each preselection criterion, in Table~\ref{tab:presel}, we give number of events for the signal with benchmark $m_a=700$ GeV and background processes after applying the preselection criteria (i)-(iv) sequentially at each stage at the HL-LHC with center-of-mass energy $\sqrt{s} =$ 14 TeV and integrated luminosity of $\mathcal{L} = 3\,\, \iab$.
Because of the huge cross section and relatively large selection efficiency, B2 dominants and accounts for $\sim$ 98\% 
of total background after the preselection.

It is worth noting that for the signals, the total preselection efficiency for $m_a = 100$ GeV is $5.96 \mltp 10^{-2}$ which is smaller than those for other masses. This is because the ALP's mass of 100 GeV is close to the mass of its decay product, the $Z$-boson, resulting in soft photons. 
One can observe from the Table~\ref{tab:presel} that the signal's efficiency passing preselection criterion (i) is only 35.6\% when $m_a =$ 100 GeV, the same efficiencies are 69.9\%, 82.9\%, 86.8\%,  89.2\% when $m_a =$ 165 Gev, 400 GeV, 1000 GeV, 2500 GeV, respectively.
As ALP's mass increases, the total preselection efficiency of signal also increases slightly, basically at $\sim 20\%-30\%$. 

\subsection{Reconstruction of the ALP mass}
\label{subsec:ALPmass}

\begin{figure}[h]
\centering
\includegraphics[width=7.2cm,height=6cm]{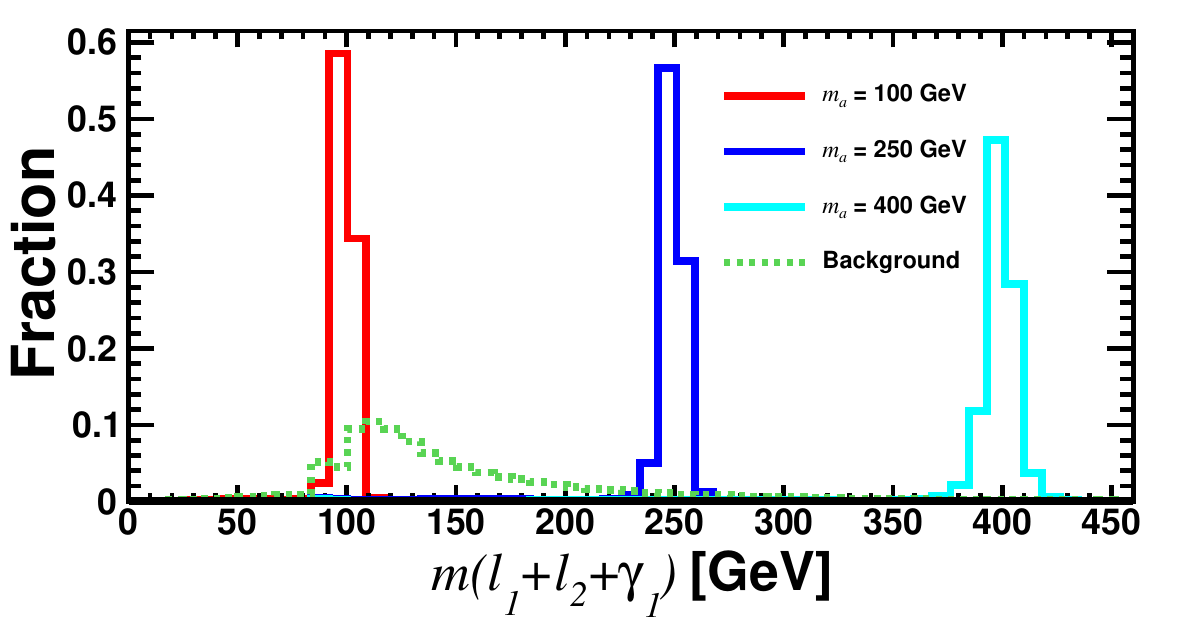}\,\,\,
\includegraphics[width=7.5cm,height=6cm]{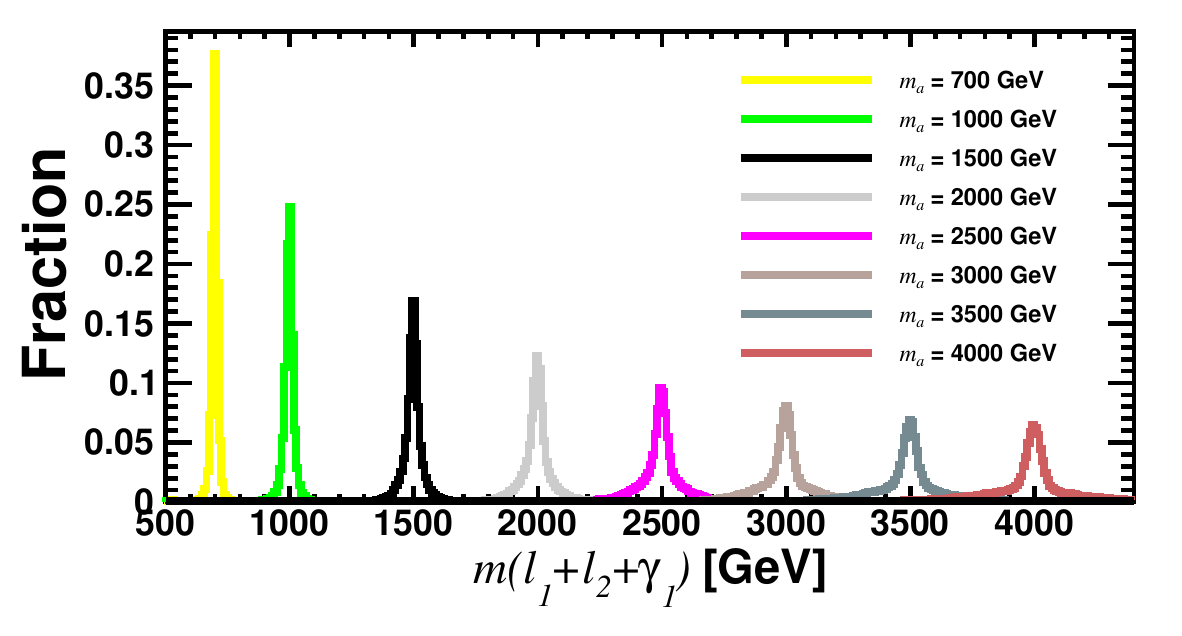}
\caption{
Distributions of invariant masses from the combined system of first two leading leptons plus photon $(\ell_1+\ell_2+\gamma_1)$ 
for the total background (dotted) and signals with assumption of various ALP masses, after the preselection at the HL-LHC with $\sqrt{s} =$ 14 TeV. 
}
\label{fig:ma1}
\end{figure}

Since the ALP decays into one photon plus one $Z$ boson, and the $Z$ boson finally decays into a pair of opposite charges leptons for the signal, the ALP mass can be reconstructed from the invariant mass of system consisted of the first leading photon and the first two leading leptons, i.e. $m(\ell_1+\ell_2+\gamma_1)$, in our analyses.
Fig.~\ref{fig:ma1} shows distributions of reconstructed ALP masses for signals with assumption of various ALP masses, after the preselection. 
One observes that reconstructed masses have sharp peaks around $m_{a}$ for signals with mass varying from 100 GeV to 4000 GeV.
Distribution for the total background (dotted) is also shown for comparison and has a wide peak around 
the background 120 GeV.
Therefore, $m(\ell_1+\ell_2+\gamma_1)$
can be used to reconstruct the ALP mass when the mass is above $\sim$ 200 GeV.
When the mass is smaller, it will suffer from large background.

\subsection{Multivariate analysis}
\label{subsec:mva}

After the preselection, to further reject the background, we input following 
thirty 
observables into 
the Toolkit for Multivariate Analysis (TMVA) 
package~\cite{Hocker:2007ht} to perform the multivariate analysis (MVA).

\begin{enumerate}[label*=\Alph*.]
\item 
Energy, transverse momentum, pseudorapidity, azimuthal angle 
of the final state objects: 
$E(O)$, $p_{_{\rm T}}(O)$, $\eta(O)$, $\varphi(O)$, with
$O = \gamma_1,\, \ell_{1},\, \ell_{2},\, j_{1},\, j_{2}$.

\item 
Magnitude and azimuthal angle of the missing transverse momentum:
$\met$, $\varphi(\met)$.

\item 
The solid angular distances ($\Delta R = \sqrt{(\Delta \eta)^2 + (\Delta \varphi)^2}$) between two objects:
$\Delta R (j_{1}, j_{2})$, $\Delta R$ ($\gamma_1$, $\ell_1+\ell_2$), 
where ``$\ell_1+\ell_2$'' stands for the combined system of the first two leptons.

\item 
The reconstructed invariant mass ($m = \sqrt{E^2 - \vec{p}^2}$) of the combined system:
$m(j_{1}+j_{2})$, $m({\ell_1+\ell_2+\gamma_1})$. 

\item 
We make all possible di-jet combinations among jets and calculate $\Delta R$ values between di-jet. $\Delta R$ and invariant mass corresponding to the di-jet combination with minimal $\Delta R$ value are input:
$\Delta R (j, j')_{\rm min}$, $m(j+j')_{ {\rm min} \Delta R}$

\item 
We make all possible di-jet combinations among jets and calculate the pseudorapidity difference $\Delta \eta$ values between di-jet. $\Delta \eta$ and invariant mass corresponding to the di-jet combination with maximal $\Delta \eta$ value are input:
$\Delta \eta(j, j')_{\rm max},$ $m(j + j')_{ {\rm max} \Delta \eta}$

\end{enumerate}

\begin{figure}[h]
\centering
\includegraphics[width=12cm,height=8cm]{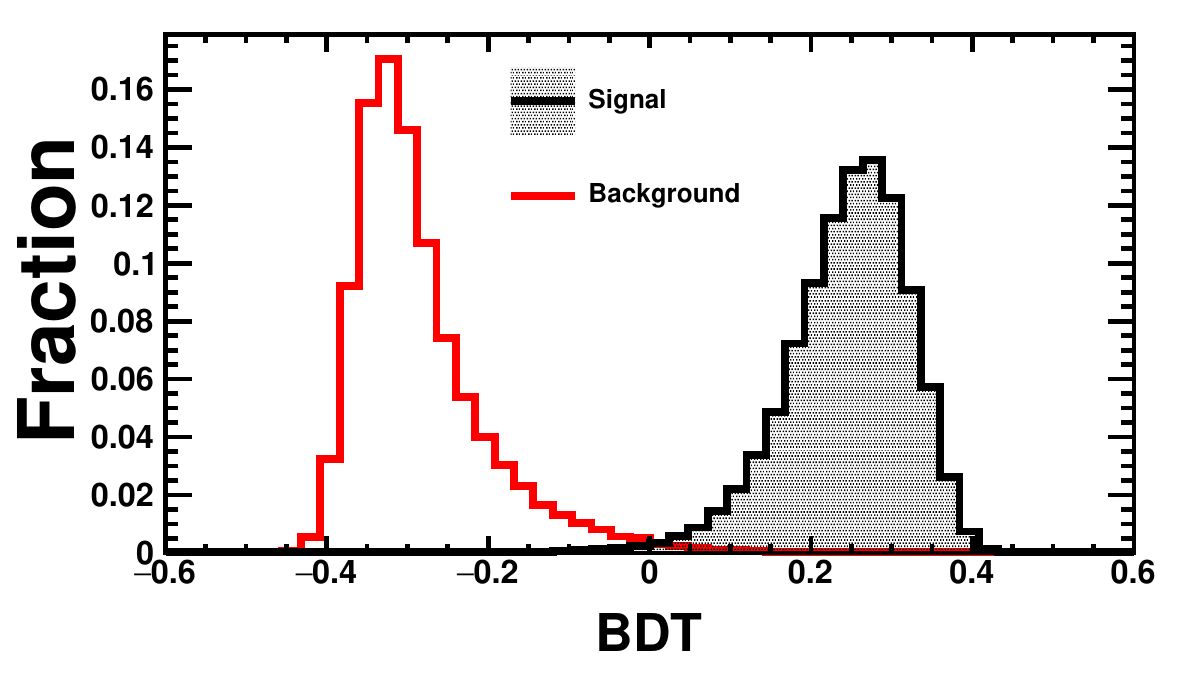}
\caption{
Distributions of BDT responses for the signal with 
benchmark ALP mass $m_{a}$ = 700 GeV (black, shaded) and total SM background (red) at the HL-LHC with $\sqrt{s} =$ 14 TeV.
}
\label{fig:BDTbenchHad}
\end{figure}

In 
Appendix~\ref{app:obs}, we show distributions of representative observables of the signal and SM background processes at the HL-LHC when assuming $m_a$ = 700 GeV.
As expected, the reconstruction mass of photons and dilepton will play a crucial role in all mass situations. 
Considering that the production process of VBF is dominant, we have added some variables into BDT that display its characteristics, for instance, the farthest two jets' pseudorapidity difference and their reconstructed invariant mass. And for the s-channel process, we have the closest two jets' solid angular distance and their reconstructed invariant mass for it. As the mass of the ALP increases, the efficiency of BDT value in distinguishing between signal and background gradually improves, and for the ALP mass $\gtrsim$ 700 GeV, the distribution of BDT values between signal and background has almost no cross regions.

In the TMVA package, the Boosted Decision Tree (BDT) algorithm with the default setting is adopted to perform the MVA and maximally reject the background.
In Fig.~\ref{fig:BDTbenchHad}, we show BDT distributions for the total background and the benchmark signal with $m_{a}$ = 700 GeV  at the HL-LHC.
Since the kinematics of signal varies with $m_a$, distributions of BDT response also change with $m_a$. 
In Appendix~\ref{app:BDT}, 
we show distributions of BDT responses of the signal and background processes at the HL-LHC with different $m_a$ assumptions. 
One observes the good separation between signal and background, and the separation becomes better as $m_a$ increases. 
This means a selection on the large value of BDT response can effectively reduce the background, especially for the heavy ALP case.
In this study, for each $m_a$ case, based on BDT distributions of the signal and total background, the BDT cut is optimized to maximal the signal statistical significance described by the following formula~\cite{cowan2012discovery, ATLAS:2020yaz, Bhattiprolu:2020mwi}:
\begin{equation}
\sigma_{\rm stat} = \sqrt{2 \left[ \left( N_s+N_b \right) \, {\rm ln}\left(1+\frac{N_s}{N_b} \right) - N_s \right] }\,\, ,
\label{eqn:statSgf}
\end{equation}
where $N_s$ ($N_b$) is the number of signal (total background) events after all selections.
In the table of 
Appendix~\ref{app:efficiency}, we present selection efficiencies of BDT cuts for both signal and background processes at the HL-LHC with different $m_a$ assumptions.

\section{Results}
\label{sec:results}

In this section, based on former analyses, we show 
discovery sensitivities on the signal at the HL-LHC with center-of-mass energy of 14 TeV and integrated luminosity of 3 $\iab$. 
Results are presented for the ALP mass $m_a$ in the range of 100 to 4000 GeV both for the specified coupling $g_{_{aWW}}$ and for the signal production cross section.

\begin{figure}[h]   
\centering
\includegraphics[width=12cm,height=8cm]{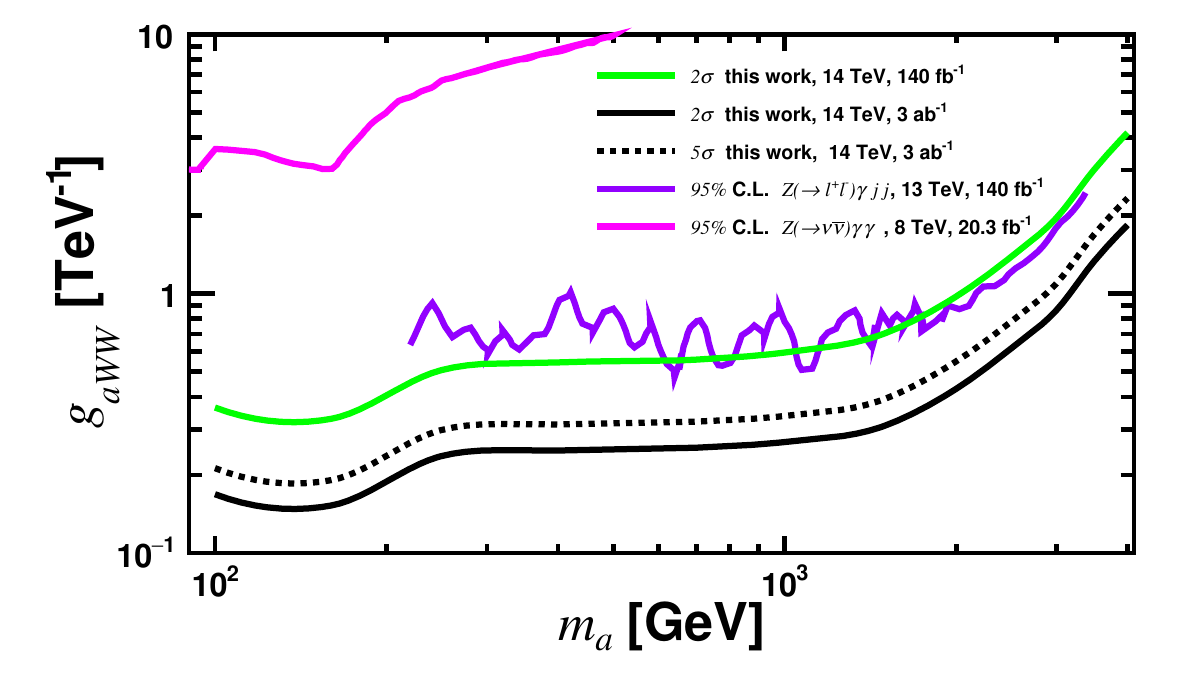}   
\caption{
Sensitivities on the coupling $g_{_{aWW}}$ as $m_a$ changes from 100 to 4000 GeV for the heavy photophobic  ($g_{a \gamma \gamma} = 0$) ALP at the HL-LHC, based on our analyses.
The red solid (dashed) curve corresponds to  $\sqrt{s} =$ 14 TeV and $\mathcal{L} = 3\,\, \iab$ at 2$\sigma$ (5$\sigma$) significance, while the green solid curve corresponds to 14 TeV, $140\,\, \ifb$ at 2$\sigma$ significance.
The 95\% C.L. results based on signal processes of  
$pp \to Z (\to \nu \bar{\nu})\, \gamma \gamma$ 
with 8 TeV and 20.3 fb$^{-1}$ (pink solid curve)~\cite{Craig:2018kne} 
and
$pp \to Z (\to \ell^{+}\ell^{-})\, \gamma j j$ ($\ell = e, \mu$) 
with 13 TeV and $140\,\, \ifb$ (purple solid curve)~\cite{Aiko:2024xiv}
are displayed in the plot for comparison.
}
\label{fig:sensALP}
\end{figure}

Under the photophobic ALP model assuming $ g_{a\gamma\gamma} =0 $, 
Fig.~\ref{fig:sensALP} presents sensitivities on the coupling $g_{_{aWW}}$ as $m_a$ changes from 100 to 4000 GeV at the HL-LHC with center-of-mass energy $\sqrt{s} =$ 14 TeV and integrated luminosities $\mathcal{L} = 3\,\, \iab$ and $140\,\, \ifb$, based on our analyses.
Results are shown at both 2 and 5$\sigma$ significances for $3\,\, \iab$ and at 2$\sigma$ significance for $140\,\, \ifb$.

With $\mathcal{L} =$ 3 $\iab$, sensitivity on $g_{_{aWW}}$ reaches the lowest value to 0.15 (0.19) TeV$^{-1}$ 
at 2-$\sigma$ (5-$\sigma$) significance when $m_a \sim$ 160 GeV. 
This is because that when $m_a$ is smaller, for example 100 GeV, although the production cross section is larger than that of 160 GeV, there are following two reasons leading to weak results: (i) the signal's preselection efficiency is low (c.f. Table~\ref{tab:preselection}) due to the soft decay products of ALP; (ii) kinematics are similar between the signal and background (c.f. Figs.~\ref{fig:ma1} and Appendix~\ref{app:BDT}), making it difficult to reject the background.
Besides, rapid decrease in the production cross section leads to weak sensitivity when $m_a$ is larger than 160 GeV. 
The 2-$\sigma$ sensitivities are nearly flat around 0.26 TeV$^{-1}$ 
as $m_a$ changes from 250 GeV to 1500 GeV.
This is mainly because that the improvement in the BDT's discrimination power offsets the decrease in the signal's production cross section.
The 2-$\sigma$ sensitivities increase rapidly to 1.8 TeV$^{-1}$ 
as $m_a$ changes from 1500 GeV to 4000 GeV. 
This is the result of small signal's production cross section. 

To compare with previous studies, 
we derive results with the same assumption at the 95\% confidence level from Refs.~\cite{Craig:2018kne} and \cite{Aiko:2024xiv} and display them in Fig.~\ref{fig:sensALP} as well. 
Ref.~\cite{Craig:2018kne} reinterprets the ATLAS experimental result~\cite{ATLAS:2016qjc} which 
considers the signal process of $pp \to Z (\to \nu \bar{\nu})\, \gamma \gamma$ 
, and the result corresponds to $\sqrt{s} =$ 8 TeV and $\mathcal{L} =$ 20.3 fb$^{-1}$.
Ref.~\cite{Aiko:2024xiv} reinterprets the ATLAS experimental result~\cite{ATLAS:2023wqy} which considers the signal process of $pp \to Z (\to \ell^{+}\ell^{-})\, \gamma j j$ with $\ell = e, \mu$
, and the result corresponds to $\sqrt{s} =$ 13 TeV and $\mathcal{L} =$ $140\,\, \ifb$. 
One observes that sensitivity result of this work are much stronger than that from Ref.~\cite{Craig:2018kne}, and with the same integrated luminosity of 140 fb$^{-1}$,  our sensitivity is similar as that from Ref.~\cite{Aiko:2024xiv} for $m_a$ in the range between 200 GeV and 3500 GeV.

\begin{figure}[h]   
\centering
\includegraphics[width=12cm,height=8cm]{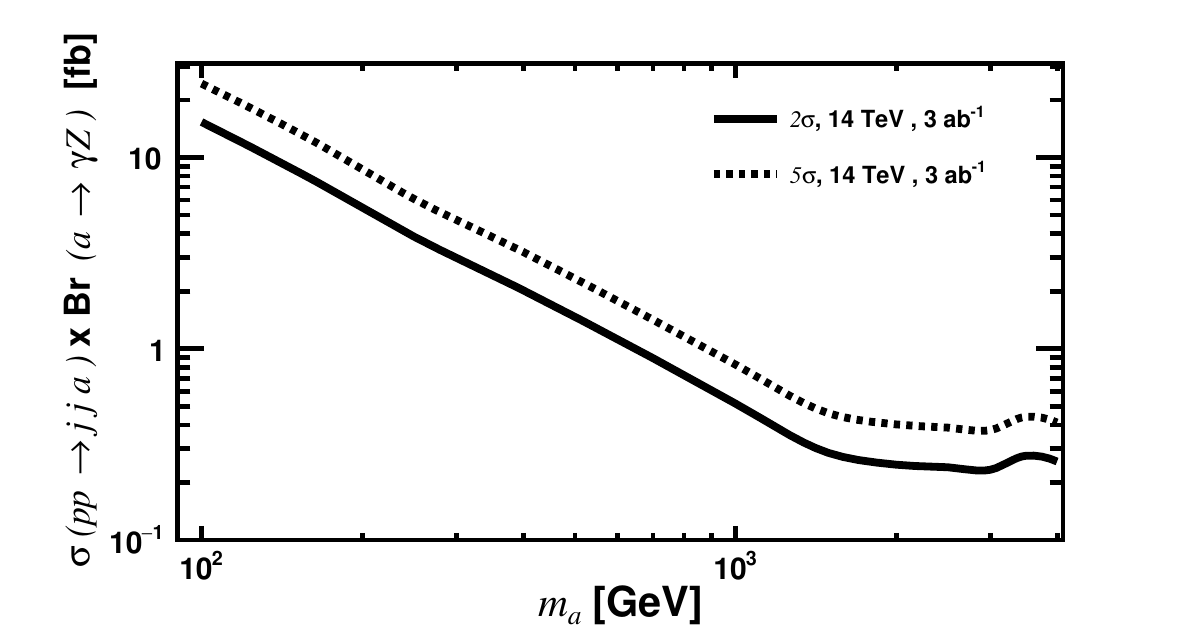} 
\caption{
Discovery sensitivities on the production cross section of the signal $\sigma(pp \to j j\, a)$ times the branching ratio Br$(a \to \gamma Z)$ as $m_a$ varies from 100 to 4000 GeV at the HL-LHC with $\sqrt{s} =$ 14 TeV and $\mathcal{L} =$ 3 ab$^{-1}$. 
The solid and dashed curves correspond to 2-$\sigma$ and 5-$\sigma$ significances, respectively. 
}
\label{fig:sensGeneral}
\end{figure}

Fig.~\ref{fig:sensGeneral} shows discovery sensitivities on the production cross section of the signal $pp \to j j\, a (\to \gamma Z)$ as varying the ALP's mass at the HL-LHC with the center-of-mass energy  of 14 TeV and integrated luminosity of 3 ab$^{-1}$, where the solid and dashed curves correspond to 2-$\sigma$ and 5-$\sigma$ significances, respectively. 
One observes that the HL-LHC experiment is sensitive to smaller production cross section for large $m_a$ cases . 
This is mainly because that the total background is smaller after the pre- and BDT selections when ALP is heavy. 
This sensitivity result can be interpreted to other theory models which have similar signal topology as in Fig.~\ref{fig:signal}.

\section{Conclusion}
\label{sec:conclusion}

Current experiments mainly probe ALPs coupling to photons and have placed strong bounds on the $g_{a\gamma\gamma}$ coupling.
This may suggest that, if ALP exists, its coupling to diphoton is suppressed. 
Therefore, it is significant to search for such heavy photophobic ALPs at colliders. 
In this article, we assume $g_{a\gamma\gamma} = 0$ and accomplish a detailed study at the HL-LHC. 
Heavy photophobic ALPs are considered to be produced associating with two jets via the s-channel exchange of vector bosons and vector boson fusions. 
The ALP then decays into a photon plus a $Z$-boson. 
To suppress the multi-jet background at the pp collider, the final state $Z$-boson is required to decay into a pair of electron or muons. 
Thus, the signal process is $pp \to j j\, a (\to  \gamma \, Z (\to \ell^{+}l^{-}) ) $ with $\ell = e, \mu$, and final state contains one photon, two oppositely charged leptons, and at least two jets. 
We consider six relevant SM background processes: $Z W \gamma$, $Z(\to l^+l^-)\, \gamma j j$, $Z Z \gamma$, $W^+ W^- \gamma$, $W(\to l\nu)\, \gamma jj$, $t \bar{t} \gamma$.

To achieve realistic data as the experiment, signal and background events are generated at center-of-mass energy of 14 TeV using the programs of MadGraph5 and PYTHIA8, and Delphes is used to perform the detector simulation. 
Branching ratios of $a \to \gamma Z, W^+ W^-, ZZ$ and  production cross section of signal process  $pp \to j j\, a (\to  \gamma Z)$ are calculated numerically using the MadGraph5 program and presented. 
For data analyses, we firstly apply preselection criteria to select the specified final state including at least one photon, exactly two oppositely charged electrons or muons, and at least two regular non-$b$ tagged jets.
Preselection efficiencies for both signals with representative ALP masses and background processes are presented. 
The ALP mass is reconstructed from the invariant mass of system consisted of the first leading photon and the first two leading leptons, i.e. $m(\ell_1+\ell_2+\gamma_1)$, and corresponding distributions for signals with assumption of various ALP masses are presented.
Thirty observables are then input and the BDT algorithm is adopted to perform the MVA and maximally reject the background.
In Appendix~\ref{app:obs}, we show distributions of representative observables of the signal and SM background processes at the HL-LHC when assuming $m_a$ = 700 GeV.
Distributions of BDT responses of the signal and background processes at the HL-LHC with different $m_a$ assumptions are shown in Appendix~\ref{app:BDT}.
The BDT cut is optimized to maximal the signal statistical significance, and selection efficiencies of BDT cuts for both signal and background processes at the HL-LHC with different $m_a$ assumptions are presented in Appendix~\ref{app:efficiency}.

Based on our analyses, sensitivities on the coupling $g_{_{aWW}}$ as $m_a$ changes from 100 to 4000 GeV are presented at the HL-LHC with center-of-mass energy $\sqrt{s} =$ 14 TeV and integrated luminosities $\mathcal{L} =$ 3 $\iab$ and $140\,\, \ifb$. 
The sensitivity on $g_{_{aWW}}$ reaches the lowest value to 0.15 (0.19) TeV$^{-1}$ 
at 2-$\sigma$ (5-$\sigma$) significance when $m_a \sim$ 160 GeV and $\mathcal{L} =$ 3 $\iab$.
The 2-$\sigma$ sensitivities are nearly flat around 0.26 TeV$^{-1}$ 
as $m_a$ changes from 250 GeV to 1500 GeV, and increase rapidly to 1.8 TeV$^{-1}$ 
as $m_a$ changes from 1500 GeV to 4000 GeV.
Results from previous studies are displayed in the same plot for comparison.
Discovery sensitivities on the production cross section of the signal $\sigma(pp \to j j\, a)$ times the branching ratio Br$(a \to \gamma Z)$ are also presented and can be interpreted to other theory models which have similar signal topology.

\appendix

\section{Effects of radiative corrections}
\label{app:RGrunning}	 

To evaluate the RG running effects, we follow most of the calculations in Ref. \cite{Bauer:2020jbp}. 
We begin from the effective Lagrangian (c.f. Eq.~(\ref{eq:L}))~\footnote{
Note that the expression of our Lagrangian is similar to that in Ref.~\cite{Aiko:2023trb} and a bit different from that in Ref. \cite{Bauer:2020jbp}.
},

\begin{equation}
\mathcal{L}_{\rm ALP} \supset
- \frac{c_{\widetilde{W}}}{f_{a}}\, a\, W_{\mu\nu}^{b}\, \widetilde{W}^{b, \mu\nu}
- \frac{c_{\widetilde{B}}}{f_{a}}\, a\,B_{\mu\nu}\, \widetilde{B}^{\mu\nu},
\label{eq:LRG}
\end{equation}

\subsection{$a\to\gamma\gamma$}

We set the ``photophobic'' condition (c.f. Eq.~(\ref{eqn:gaGG})), 
\begin{equation}
g_{a\gamma\gamma}
=
\frac{4}{f_a}\left(s^2_\theta c_{\widetilde{W}}+c^2_\theta c_{\widetilde{B}}\right)
=
\frac{16\pi\alpha}{f_a}\left(\frac{c_{\widetilde{W}}}{g^2}+\frac{c_{\widetilde{B}}}{g^{\prime 2}}\right)
=
0\,,
\label{eqn:photophobic}
\end{equation}
at a high energy scale $\Lambda = 4 \pi f_a$, where $\alpha = e^2 / (4 \pi)$ is the fine-structure constant for the electromagnetic interaction; 
$g$ and $g^{\prime}$ are the coupling constants of the SU(2) and U(1) gauge groups, respectively.
Since the operators $g^2W_{\mu\nu}^b\widetilde{W}^{b,\mu\nu}$ and $g^{\prime2}B_{\mu\nu}\widetilde{B}^{\mu\nu}$ do not have anomalous dimensions at one- and two-loop level~\cite{Bauer:2020jbp}, the corresponding Wilson coefficients $c_{\widetilde{W}}/g^2$ and $c_{\widetilde{B}}/g^{\prime2}$ are scale independent; or equivalently, the following conditions hold until two-loop level:
\begin{equation}
\frac{d}{d\ln\mu}\left(\frac{c_{\widetilde{W}}}{g^2}\right)=\frac{d}{d\ln\mu}\left(\frac{c_{\widetilde{B}}}{g^{\prime2}}\right)=0 \,.
\end{equation}
Thus we can straightforwardly obtain that $\frac{d}{d\ln\mu}\left(\frac{c_{\widetilde{W}}}{g^2}+\frac{c_{\widetilde{B}}}{g^{\prime 2}}\right)=0$, or equivalently, $\frac{c_{\widetilde{W}}}{g^2}+\frac{c_{\widetilde{B}}}{g^{\prime 2}}=0$ does not depend on the scale $\mu$, meaning that the ``photophobic'' condition (Eq.~(\ref{eqn:photophobic})) keeps at the whole scale until two-loop level.

The leading contribution for the $a\rightarrow\gamma\gamma$ decay comes at one-loop level from the $W$-loops without RGE running effects, whose partial decay width is shown as \cite{Bauer:2020jbp,Aiko:2023trb}
\begin{equation}
\Gamma_{a\rightarrow\gamma\gamma}=\frac{m_a^3}{64\pi}\left|g_{a\gamma\gamma}^{\textrm{eff.}}\right|^2=\frac{\alpha^2c^2_{\widetilde{W}}m_a^3}{\pi^3f_a^2}
\left|B\left(\frac{4m^2_W}{m^2_a}\right)\right|^2.
\end{equation}
Here, $g_{a\gamma\gamma}^{\textrm{eff.}}$ is the effective ALP-photon coupling after loop calculation, $m_W$ is the $W$-boson mass, and the function 
\begin{equation}
B(x)=1-(x-1)f^2(x);
\end{equation}
with
\begin{equation}
\label{eq:loop}
f(x)=\left\{\begin{array}{cc}\arcsin\left(x^{-1/2}\right),&(x\geq1);\\ \frac{\pi}{2}+\frac{\textrm{i}}{2}\ln\frac{1+\sqrt{1-x}}{1-\sqrt{1-x}},&(x<1).\end{array}\right.
\end{equation}

\subsection{$a \to f \bar{f}$}

The ALP couplings to fermions are generated through RG running effects. The decay widths are calculated as \cite{Bauer:2020jbp,Aiko:2023trb}
\begin{equation}
\Gamma_{a\rightarrow f\bar{f}|_{f=q,\ell}}=\frac{N^f_c m_a m^2_f}{8\pi f_a^2}\left|c_{ff}(m_a)\right|^2\sqrt{1-\frac{4m_f^2}{m^2_a}}\, ,
\end{equation}
where $m_f$ is the fermion mass, $N^f_c$ = 1 (3) denotes the color factor for leptons (quarks), and dependencies of the coefficients on the energy scale $\mu$ are 
\begin{eqnarray}
c_{uu}(\mu)&=&-\left(\frac{9\alpha}{4\pi s^2_\theta}c_{\widetilde{W}}+\frac{17\alpha}{12\pi c^2_\theta}c_{\widetilde{B}}\right)\ln\frac{\mu}{\Lambda};\\
c_{dd}(\mu)&=&-\left(\frac{9\alpha}{4\pi s^2_\theta}c_{\widetilde{W}}+\frac{5\alpha}{12\pi c^2_\theta}c_{\widetilde{B}}\right)\ln\frac{\mu}{\Lambda};\\
c_{\ell\ell}(\mu)&=&-\left(\frac{9\alpha}{4\pi s^2_\theta}c_{\widetilde{W}}+\frac{15\alpha}{4\pi c^2_\theta}c_{\widetilde{B}}\right)\ln\frac{\mu}{\Lambda}.
\end{eqnarray}
Note that since $m_f$ also depends significantly on $\mu$ for quarks, the RG running effects for quarks masses are included as well when we calculate the ALP's decay branching ratio to the quark pair.

\subsection{$a \to g g$}

Due to the nonzero ALP-quark couplings after RGE running, we can then calculate the $a\rightarrow gg$ decay process including a further finite loop calculation as
\begin{equation}
\Gamma_{a\rightarrow gg}=\frac{\alpha_s^2m_a^3}{8\pi^3f_a^2}\left|c_{GG}^{\textrm{eff.}}(m_a)\right|^2,
\end{equation}
where $\alpha_s=g_s^2/(4 \pi)$ denotes the strong coupling parameter, and the effective ALP-gluon coupling after loop calculation is 
\begin{equation}
c_{GG}^{\textrm{eff.}}(m_a)=\frac{1}{2}\sum_qc_{qq}(m_a)B'\left(\frac{4m_q^2}{m^2_a}\right),
\end{equation}
where the function
\begin{equation}
B'(x)=1-xf^2(x),
\end{equation}
and the expression of $f(x)$ can be found in Eq.~(\ref{eq:loop}).

\subsection{Numerical results}

\begin{figure}[h]
\includegraphics[width=7.3cm,height=5.3cm]{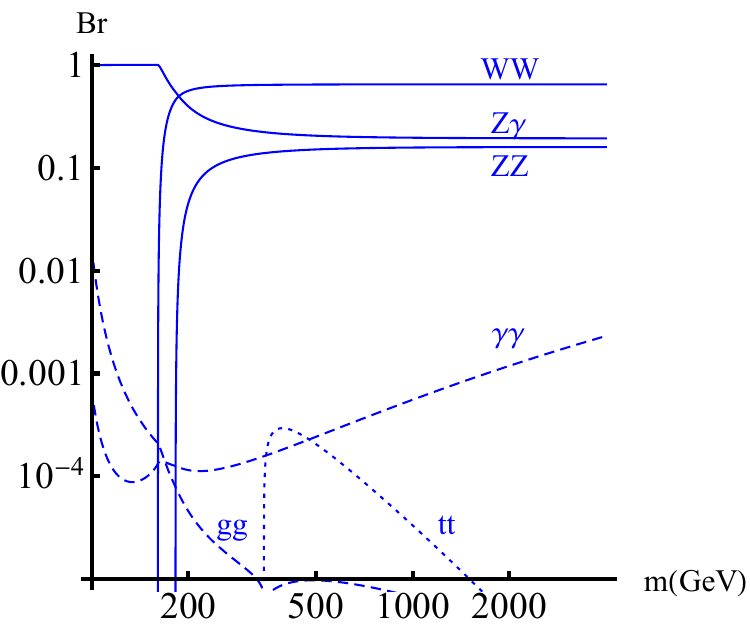}\,\,\,\,\,
\includegraphics[width=7.3cm,height=5.3cm]{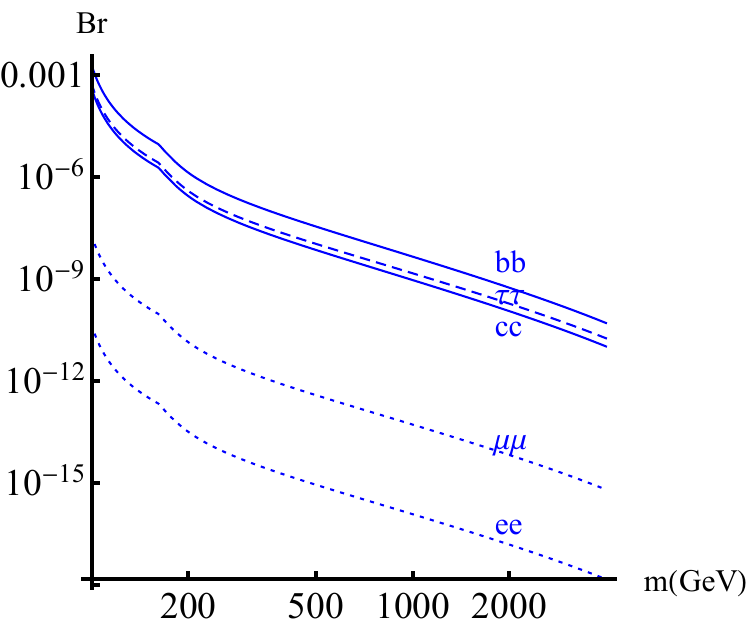}
\caption{
The decay branching ratios of the photophobic ALP for some typical channels as a function of $m_a$.
}
\label{fig:BR}
\end{figure}

Combining all these calculations at a benchmark point $f_a=1~\textrm{TeV}$ and $\Lambda=4\pi f_a$, we show the decay branching ratios of the photophobic ALP for some typical channels depending on $m_a$ in Fig. \ref{fig:BR}.
As can be seen, in the mass region $100~\textrm{GeV}<m_a<4000~\textrm{GeV}$ which we considered in this study, all these additional channels have the branching ratios $<10^{-2}$. This means that when we focus on the $a\rightarrow WW,ZZ,Z\gamma$ channels, the modifications on their branching ratios due to such effects are all ignorable.

\section{Distributions of representative observables}
\label{app:obs}	 

\begin{figure}[H]
\centering
\addtocounter{figure}{1}
\subfigure{
\includegraphics[width=7.3cm,height=5.3cm]{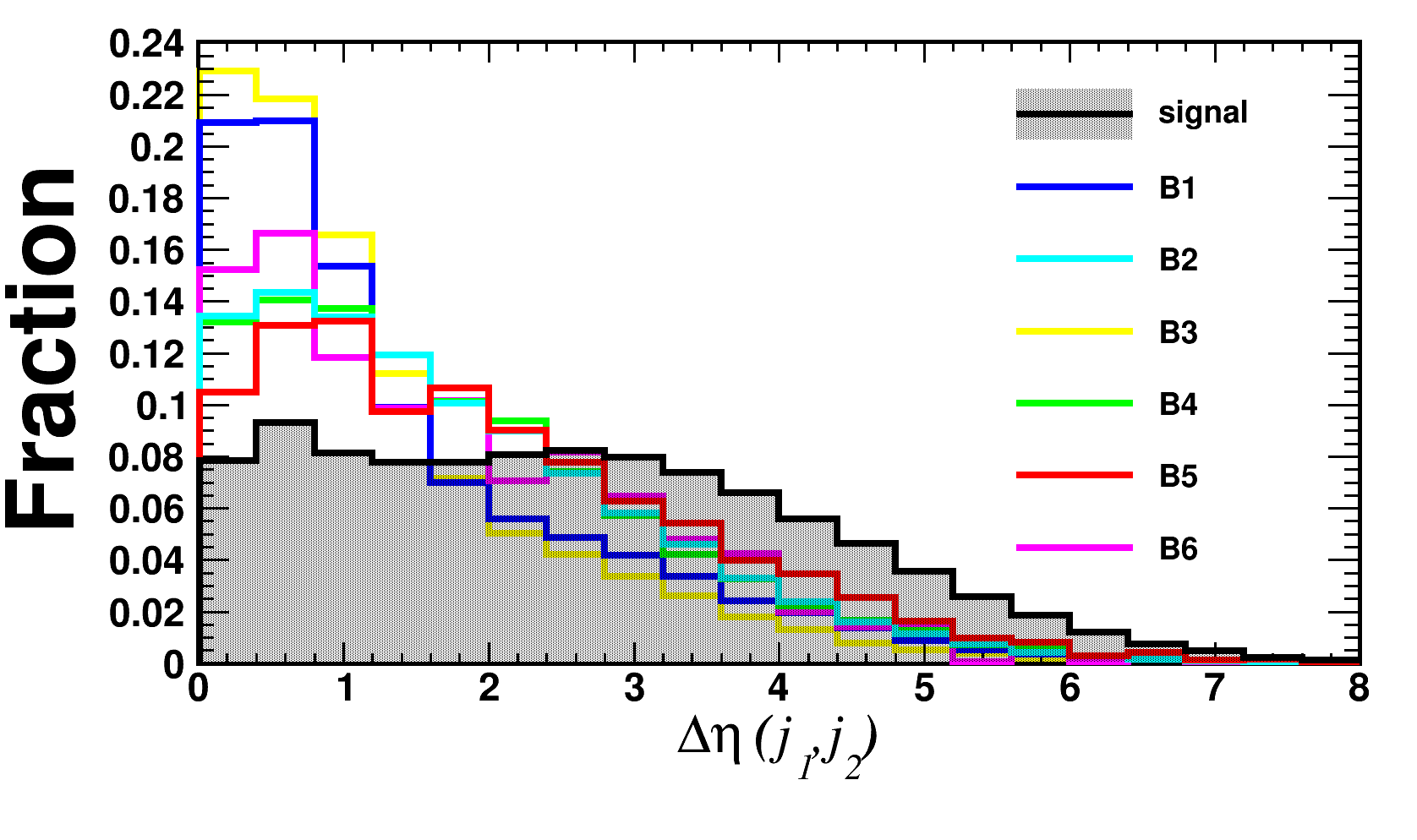}
\includegraphics[width=7.3cm,height=5.3cm]{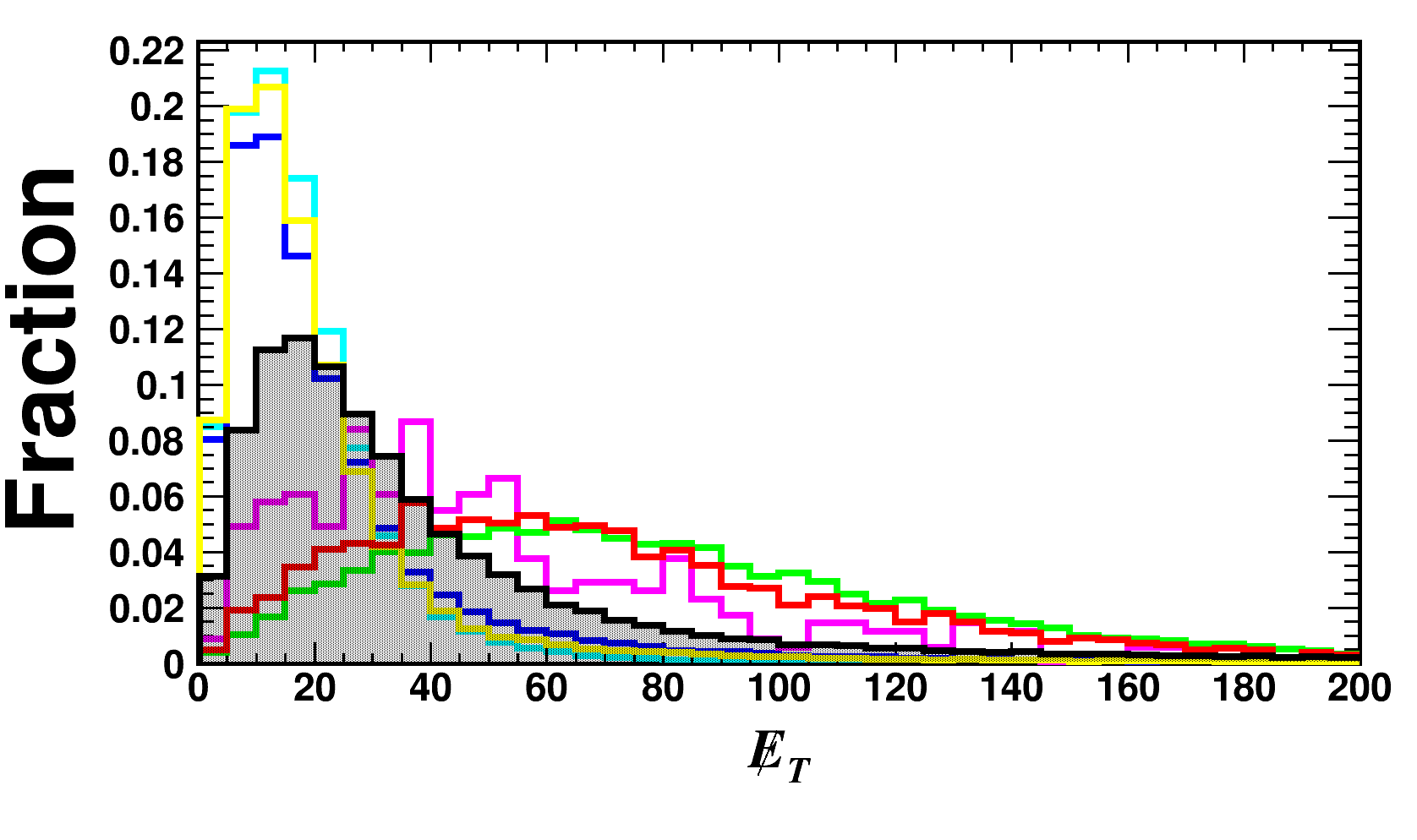}
}
\end{figure}
\addtocounter{figure}{-1}
\vspace{-1.0cm}
\begin{figure}[H]
\centering
\addtocounter{figure}{1}
\subfigure{
\includegraphics[width=7.3cm,height=5.3cm]{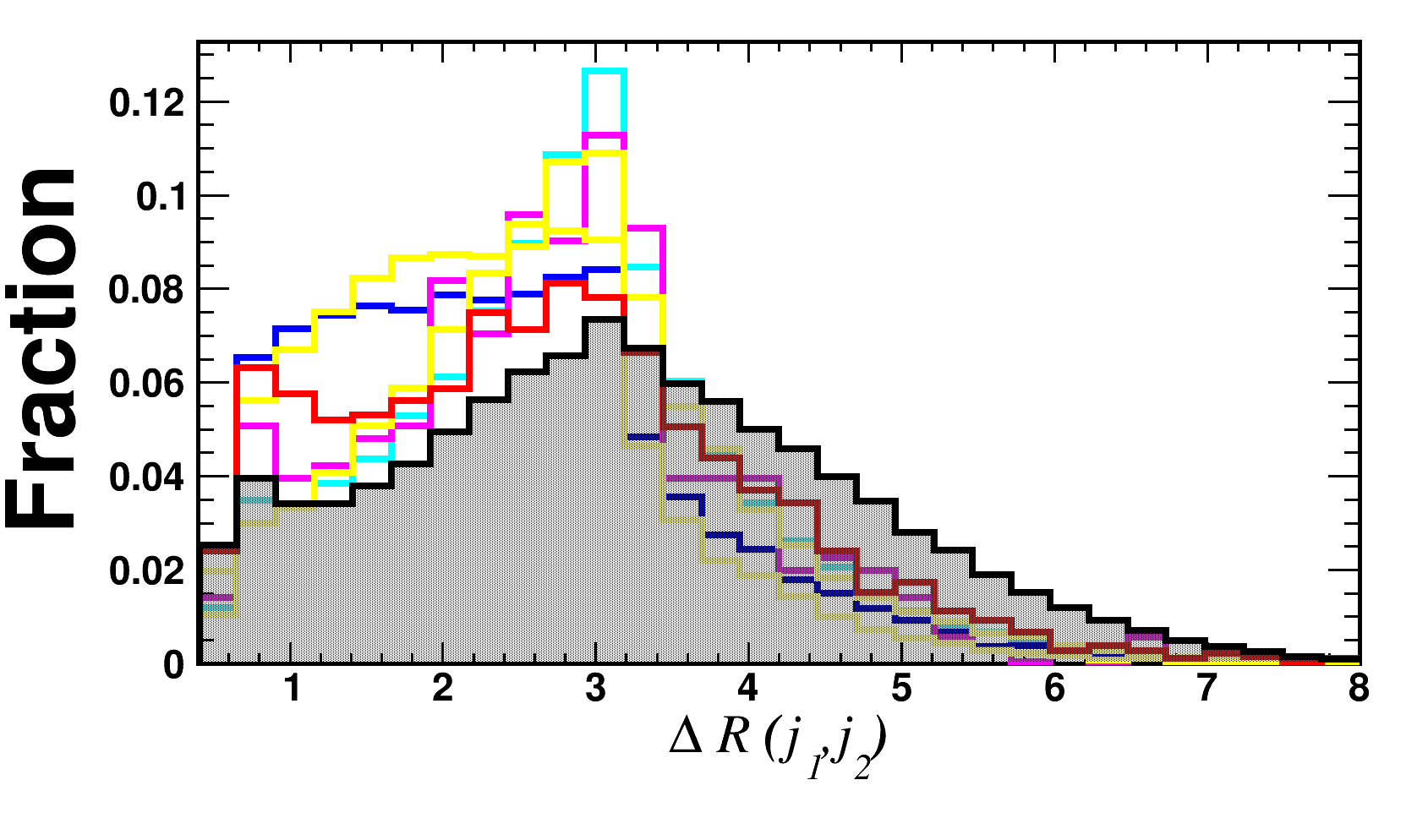}
\includegraphics[width=7.3cm,height=5.3cm]{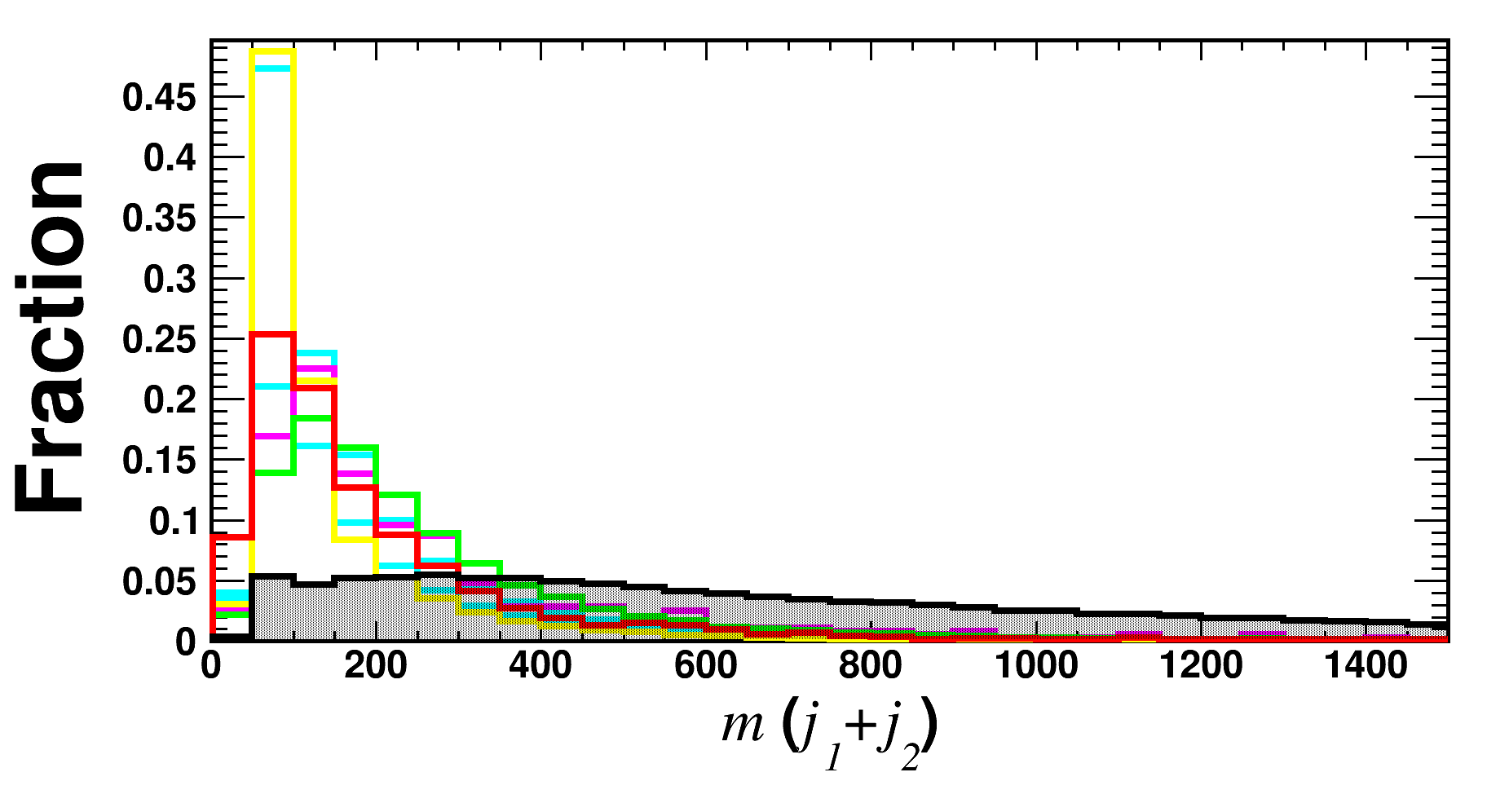}
}
\end{figure}
\vspace{-1.0cm}
\begin{figure}[H] 
\centering
\addtocounter{figure}{-1}
\subfigure{
\includegraphics[width=7.3cm,height=5.3cm]{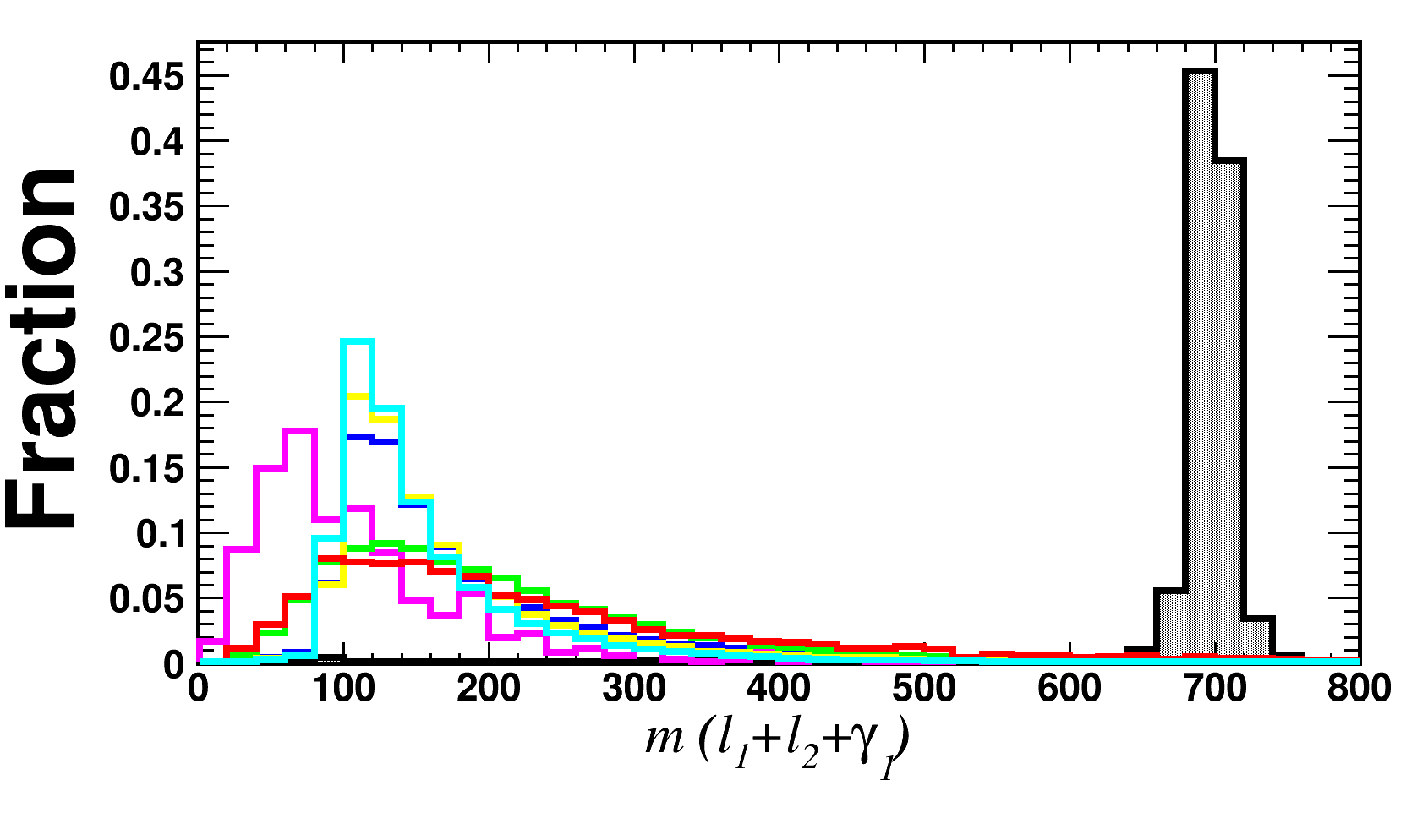}	
\includegraphics[width=7.3cm,height=5.3cm]{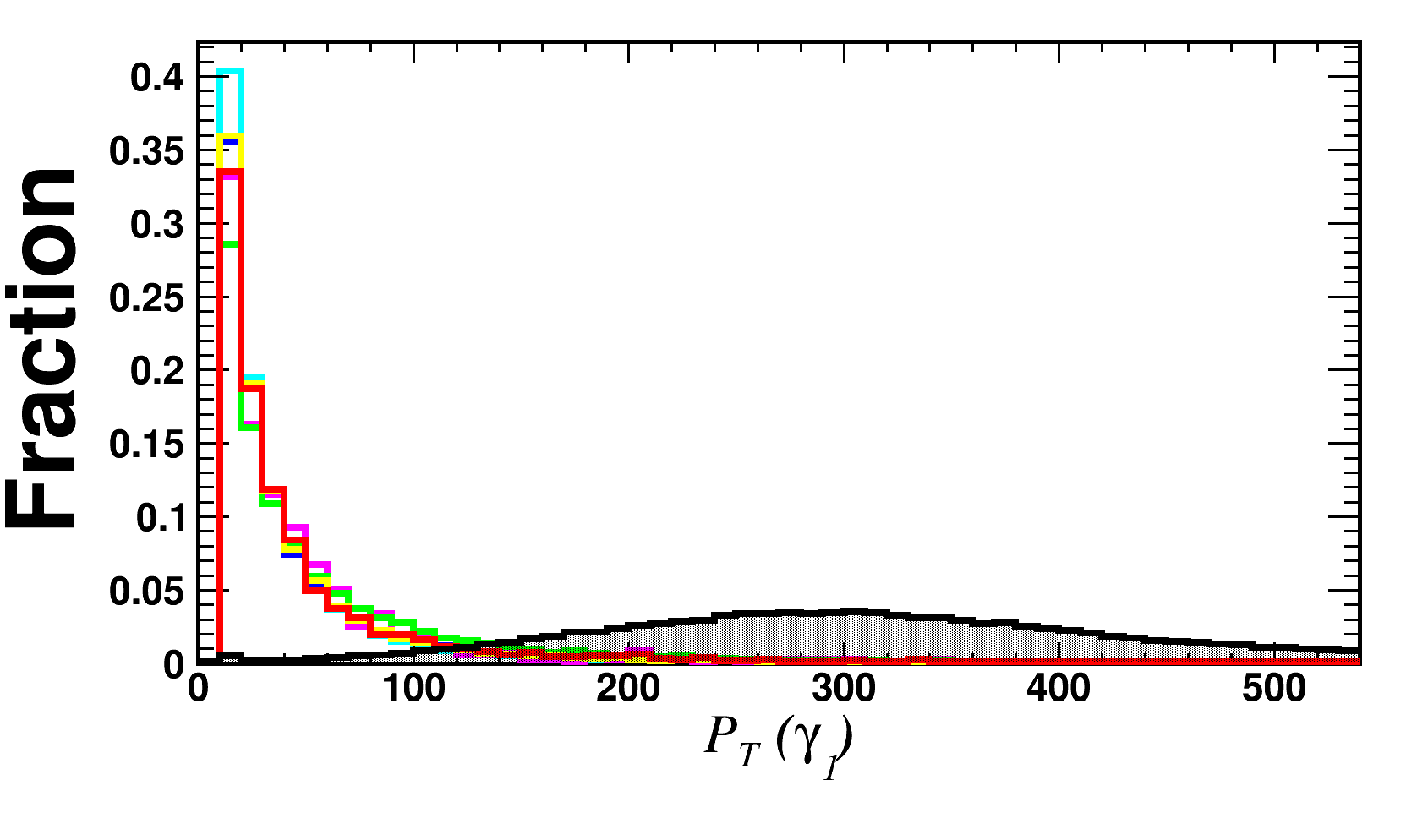}
}
\end{figure}
\vspace{-1.0cm}
\begin{figure}[H] 
\centering
\addtocounter{figure}{1}
\subfigure{
\includegraphics[width=7.3cm,height=5.3cm]{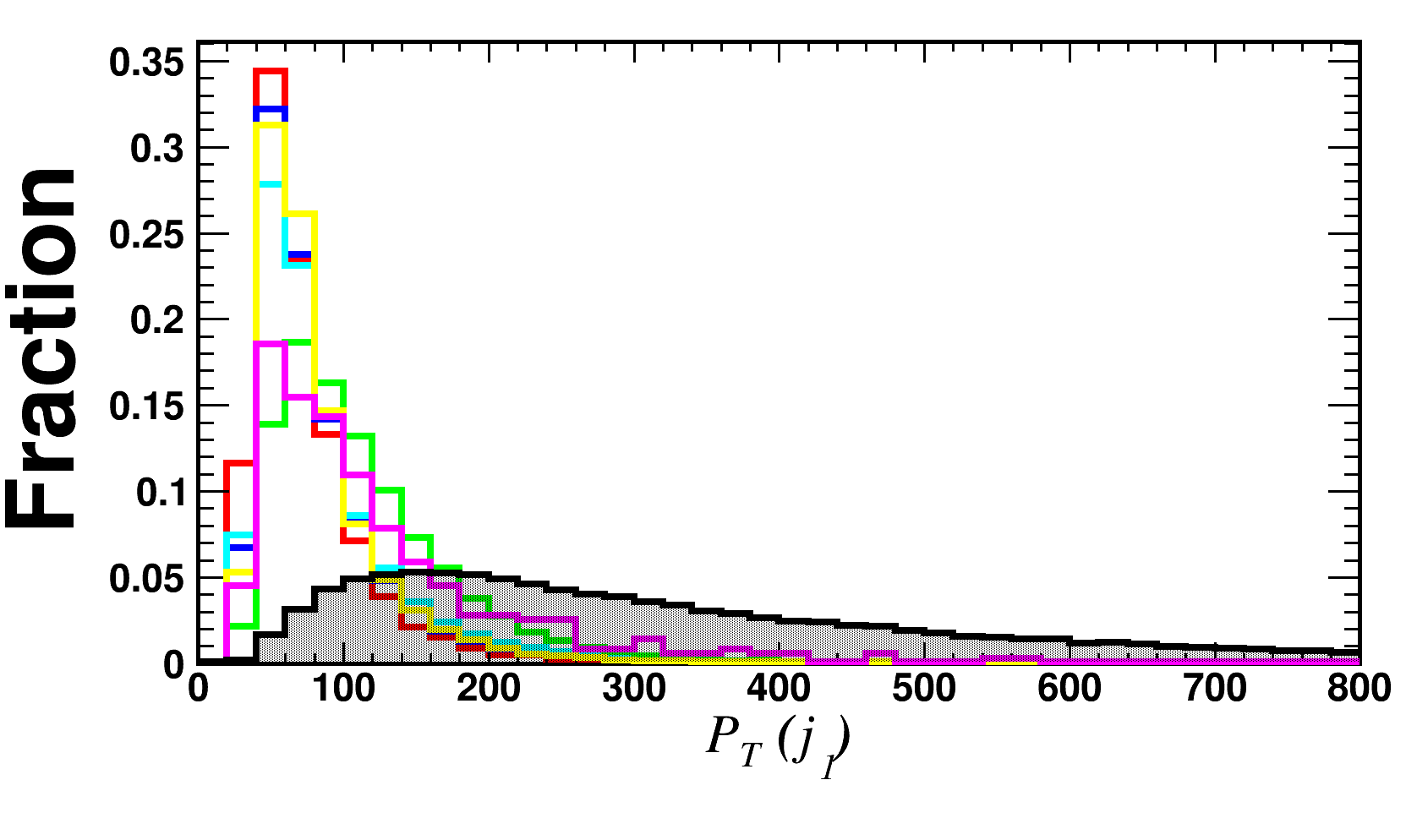}
\includegraphics[width=7.3cm,height=5.3cm]{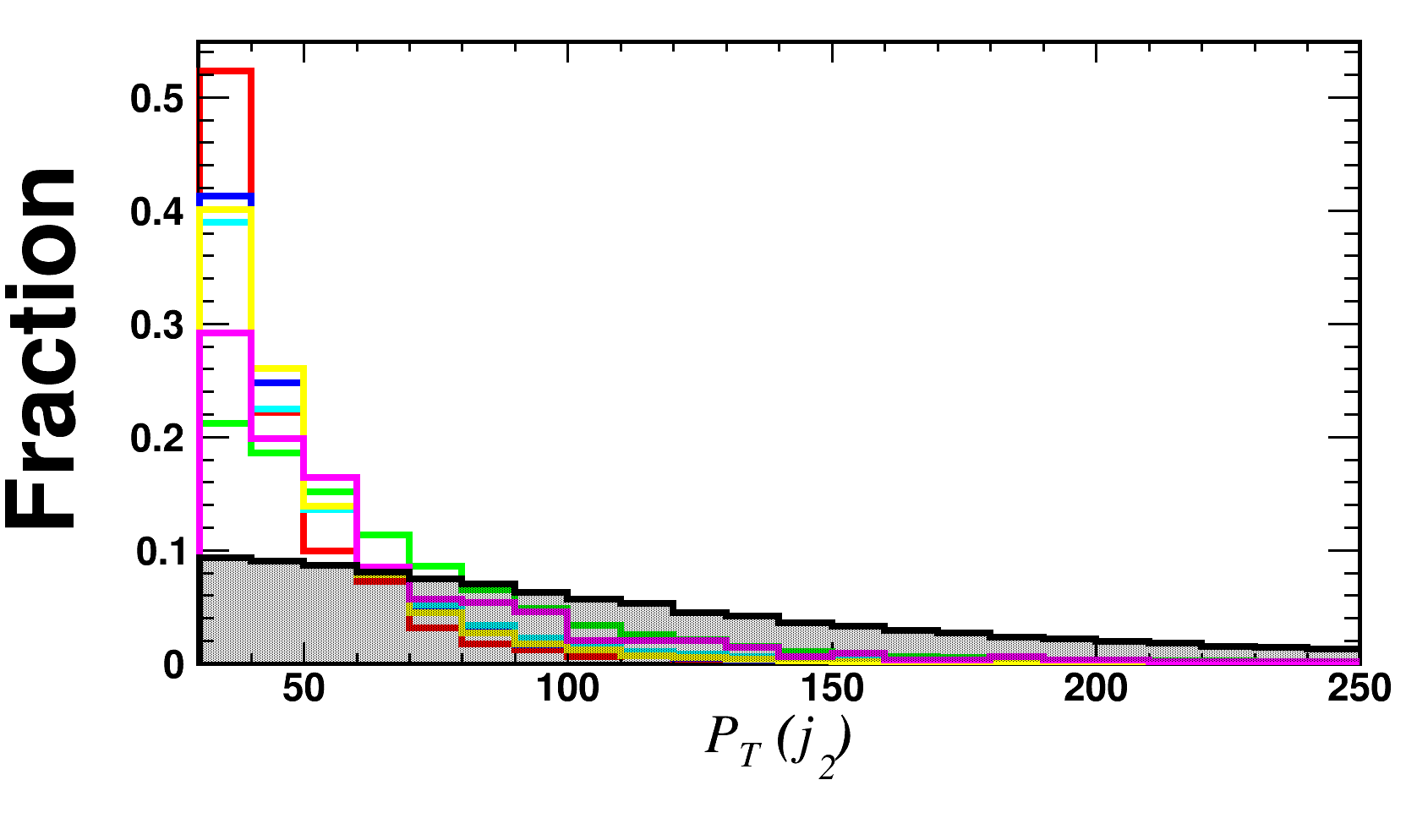}
}
\end{figure}
\vspace{-1.0cm}
\begin{figure}[H] 
\centering
\addtocounter{figure}{-1}
\subfigure{
\includegraphics[width=7.3cm,height=5.3cm]{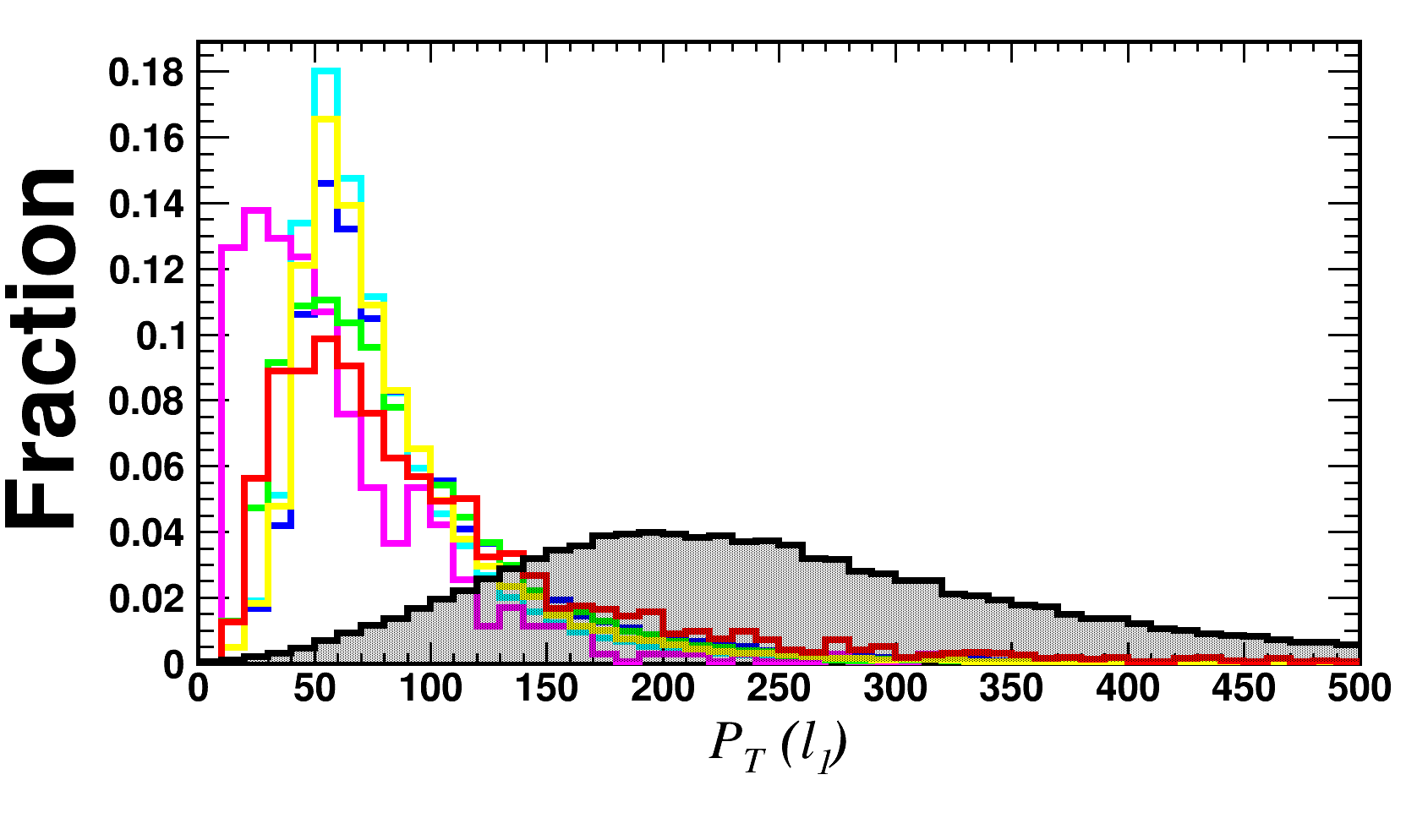}
\includegraphics[width=7.3cm,height=5.3cm]{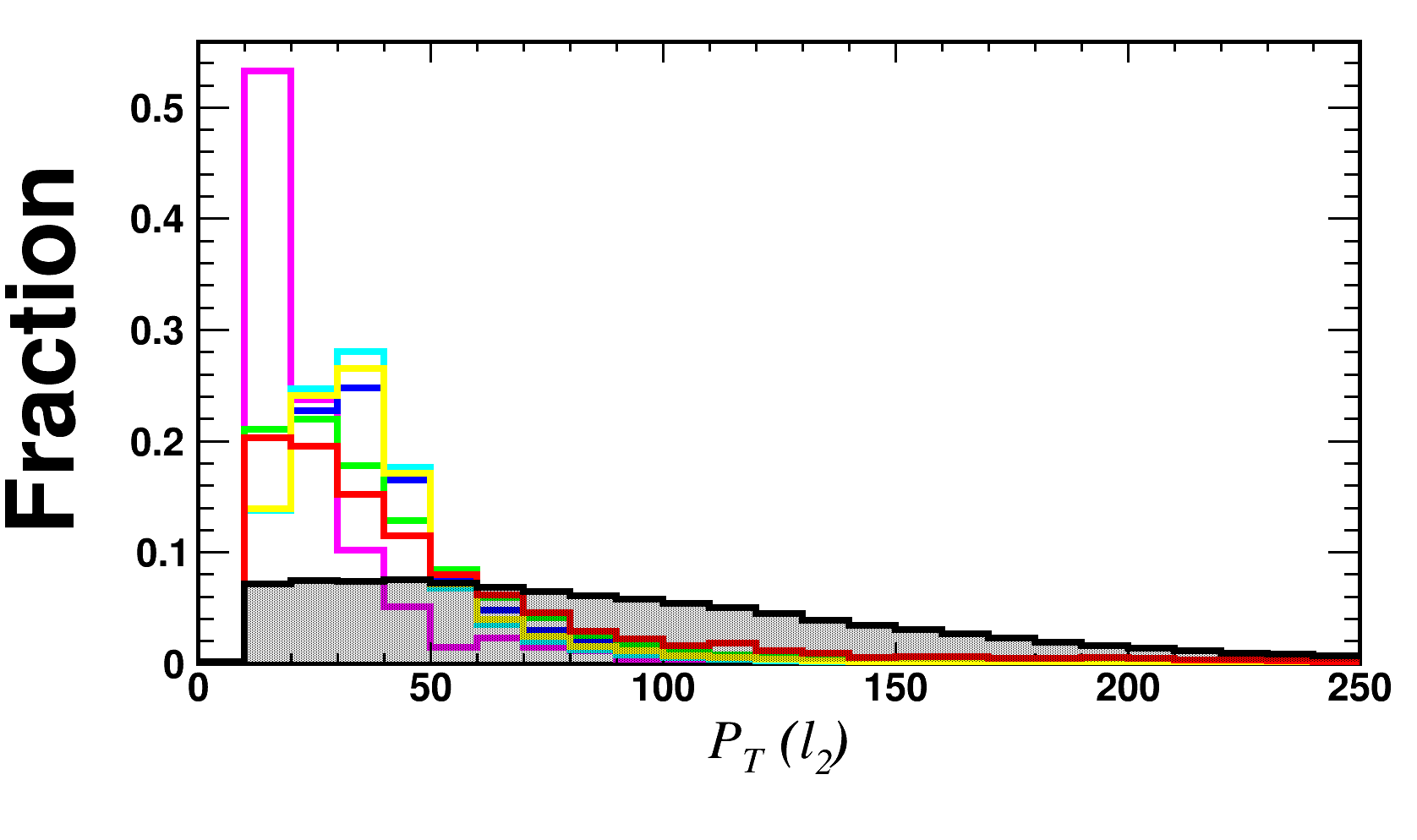}
}
\caption{
Distributions of representative observables of the signal (black, dashed) and six background processes at the HL-LHC with $\sqrt{s} =$ 14 TeV, assuming ALP mass $m_a$ = 700 GeV.
}
\label{fig:obs}
\end{figure}

\section{Distributions of BDT responses}
\label{app:BDT}

\begin{figure}[H]
\centering
\subfigure
{
\includegraphics[width=7.3cm,height=5.3cm]{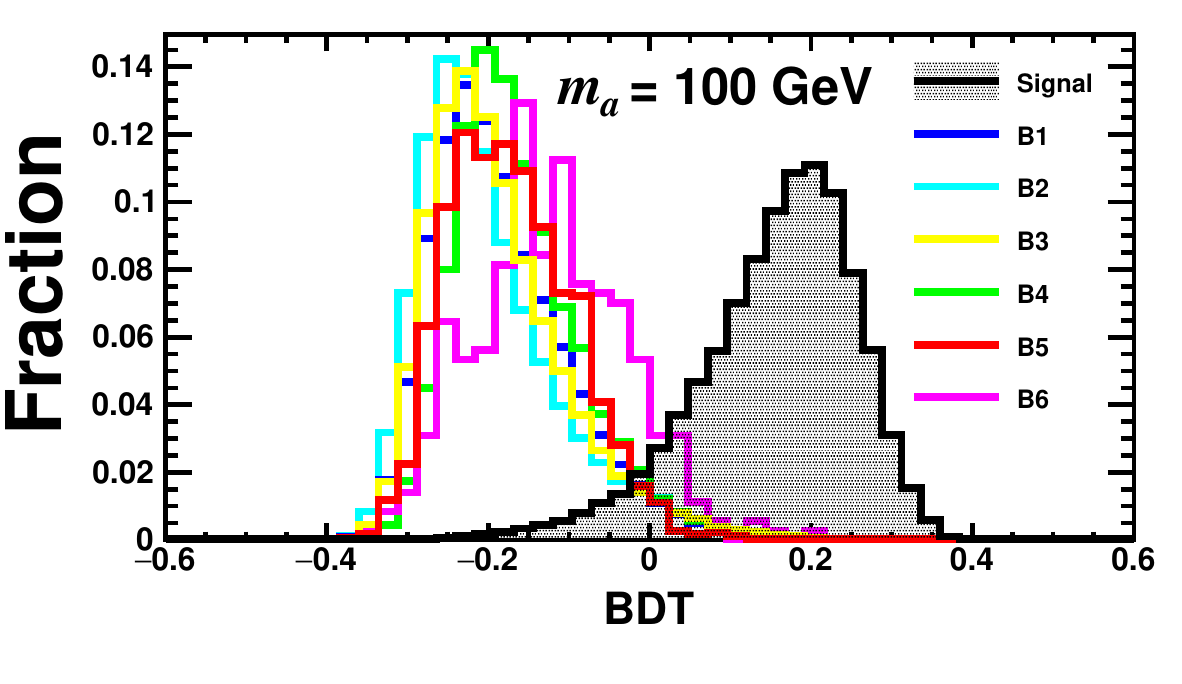}\,\,\,\,\,\,\,\,
\includegraphics[width=7.3cm,height=5.3cm]{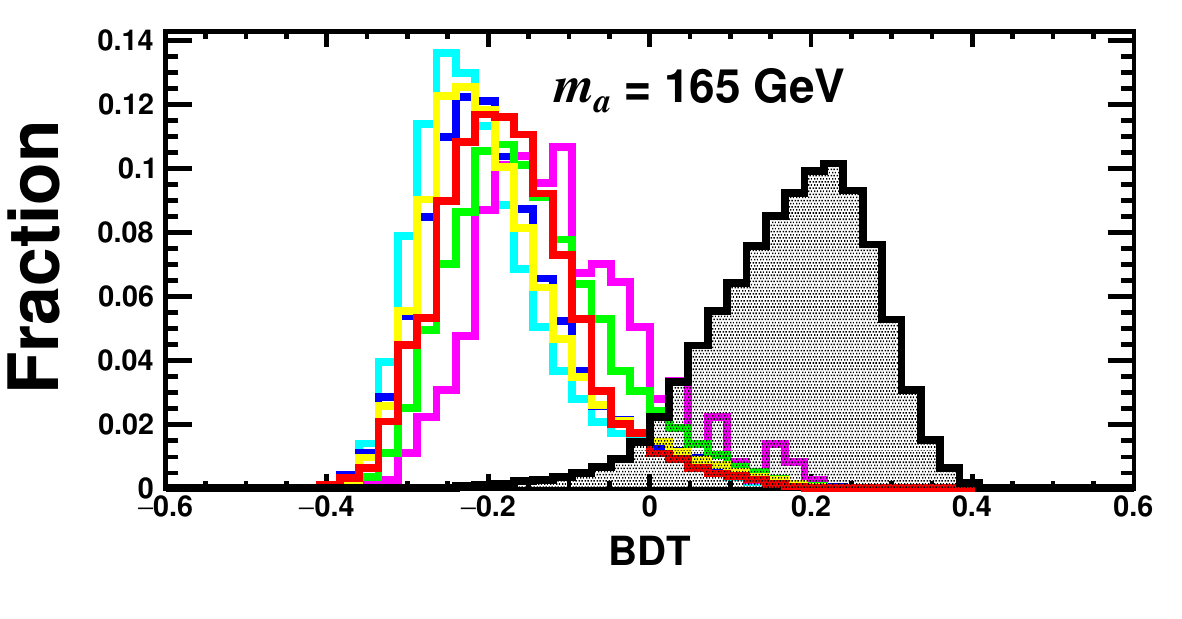}
}
\end{figure}
\addtocounter{figure}{-1}
\vspace{-1.0cm}
\begin{figure}[H]
\centering
\addtocounter{figure}{1}
\subfigure{
\includegraphics[width=7.3cm,height=5.3cm]{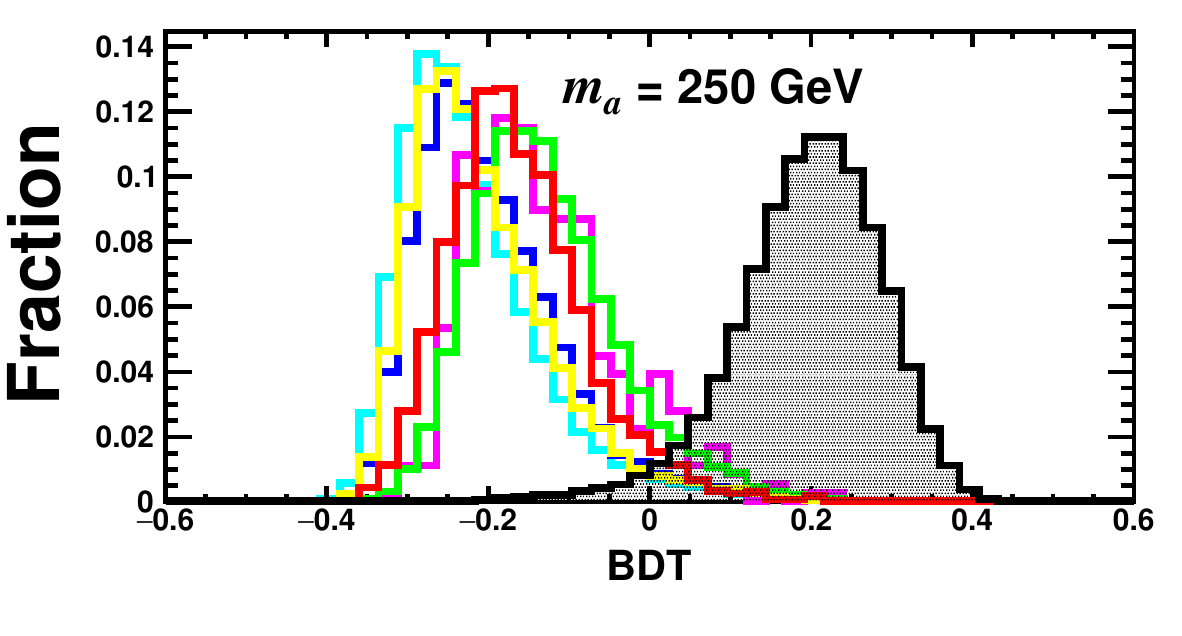}\,\,\,\,\,\,\,\,
\includegraphics[width=7.3cm,height=5.3cm]{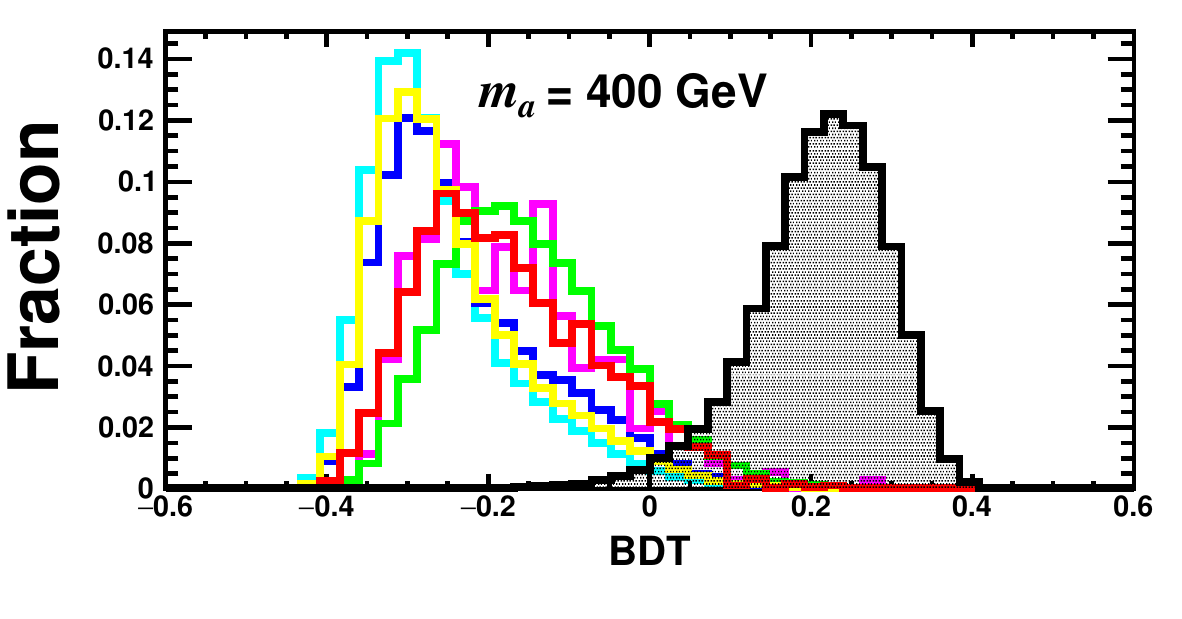}
}
\end{figure}
\vspace{-1.0cm}
\begin{figure}[H] 
\centering
\addtocounter{figure}{-1}
\subfigure{
\includegraphics[width=7.3cm,height=5.3cm]{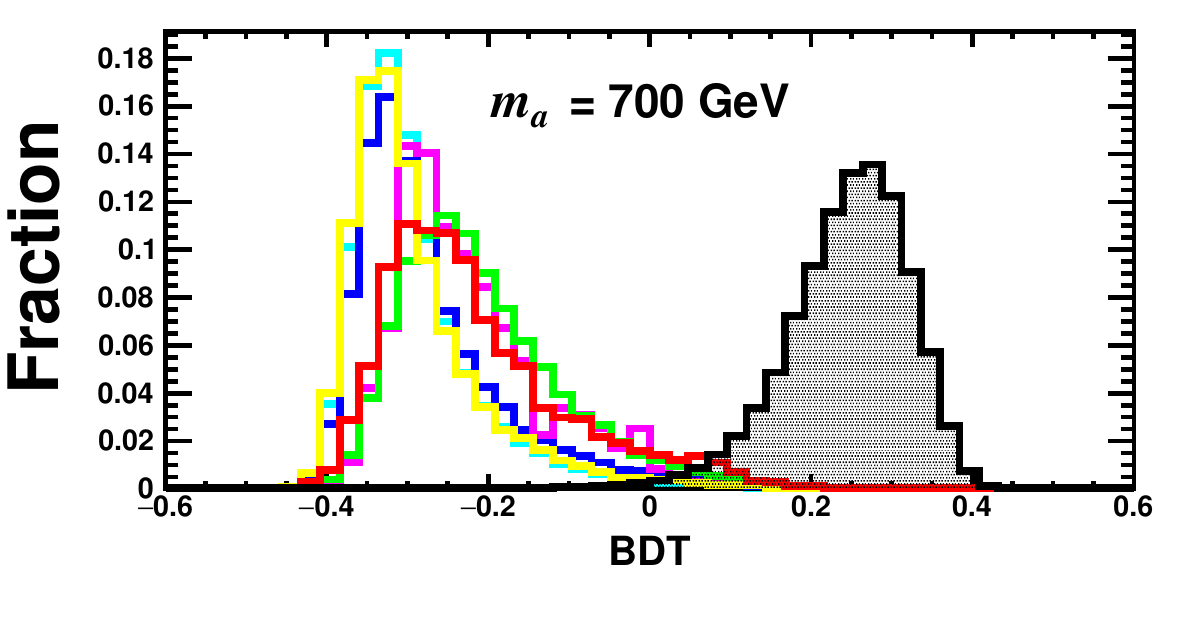}\,\,\,\,\,\,\,\,
\includegraphics[width=7.3cm,height=5.3cm]{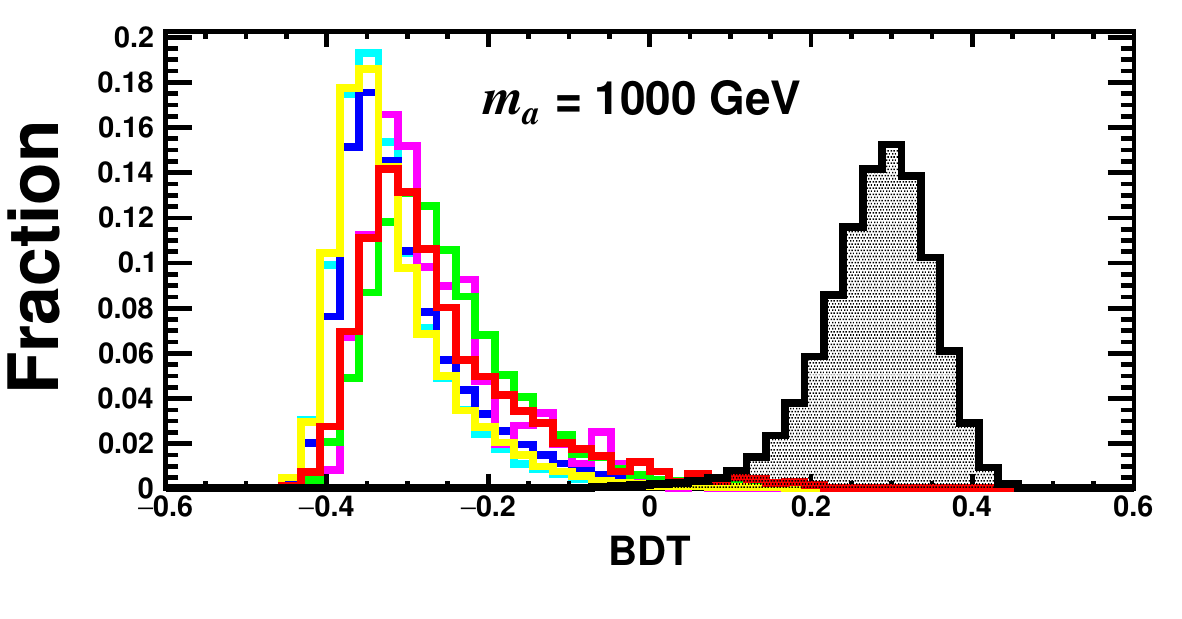}
}
\end{figure}
\vspace{-1.0cm}
\begin{figure}[H] 
\centering
\addtocounter{figure}{1}
\subfigure{
\includegraphics[width=7.3cm,height=5.3cm]{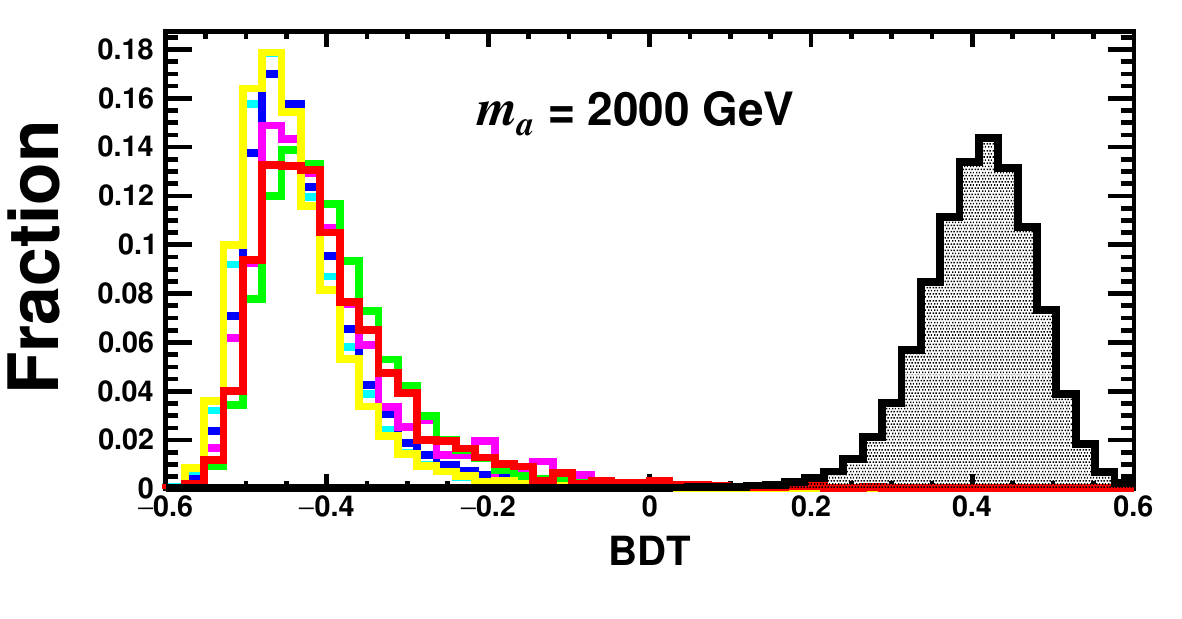}\,\,\,\,\,\,\,\,
\includegraphics[width=7.3cm,height=5.3cm]{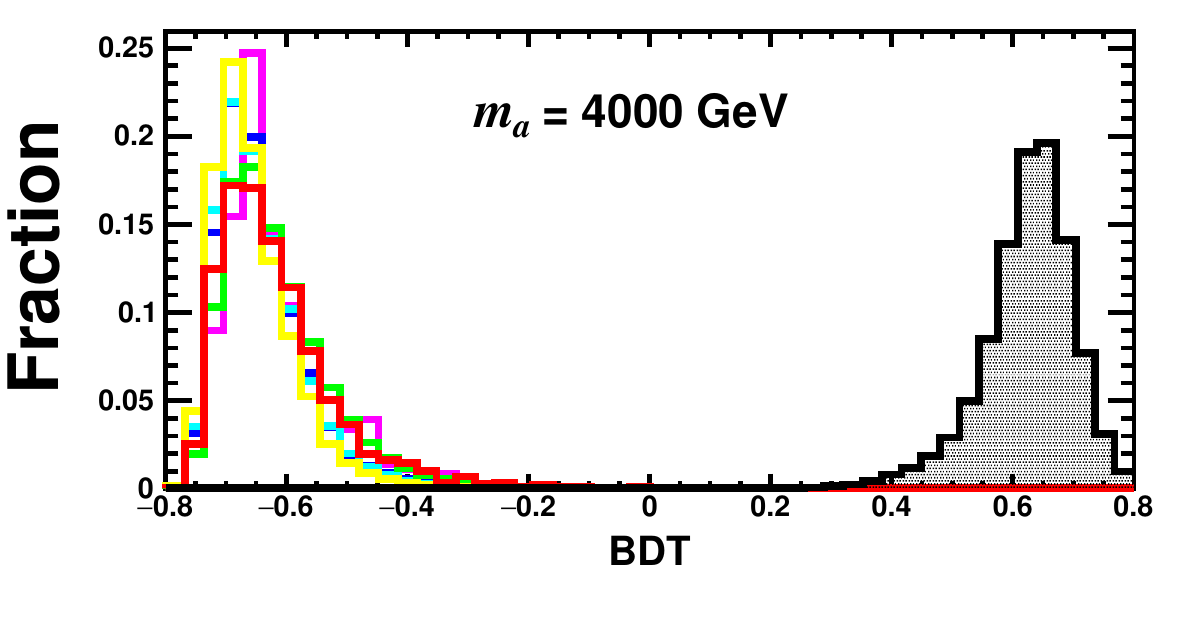}
}
\caption{
Distributions of BDT responses of the signal (black, shaded) and six background processes at the HL-LHC with $\sqrt{s} =$ 14 TeV, assuming different ALP masses.
}
\label{fig:BDTall}
\end{figure}

\section{The BDT selection efficiency table}
\label{app:efficiency}

\begin{table*}[h]
\centering 
\scalebox{0.75}{
\begin{tabular}{ccccccccc}
\hline
\hline
$m_a$ [GeV] & BDT & signal & $ Z W \gamma$ & $ Z(\to l^+l^-) \gamma j j$ & $ Z Z \gamma$& $W^+ W^- \gamma$ & $W(\to l\nu) \gamma jj$ & $ t\bar{t} \gamma $ \\
\hline
100 &  0.117  &$7.00\mltp10^{-1}$&$3.45\mltp10^{-3}$&$3.83\mltp10^{-3}$&$6.27\mltp10^{-3}$&$1.73\mltp10^{-2}$&$-$ &$1.12\mltp10^{-2}$ \\
165 & 0.129  &$7.03\mltp10^{-1}$&$6.21\mltp10^{-3}$&$4.07\mltp10^{-3}$&$8.00\mltp10^{-3}$&$6.04\mltp10^{-2}$&$2.53\mltp10^{-2}$&$5.05\mltp10^{-4}$\\
250 & 0.155  &$7.13\mltp10^{-1}$&$3.89\mltp10^{-3}$&$2.09\mltp10^{-3}$&$3.16\mltp10^{-3}$&$3.87\mltp10^{-2}$&$2.81\mltp10^{-2}$&$2.24\mltp10^{-4}$\\
400 & 0.167  &$7.32\mltp10^{-1}$&$1.44\mltp10^{-3}$&$9.20\mltp10^{-4}$&$1.53\mltp10^{-3}$&$2.65\mltp10^{-2}$&$2.53\mltp10^{-2}$&$5.61\mltp10^{-5}$\\
700 & 0.194  &$7.75\mltp10^{-1}$&$6.41\mltp10^{-5}$&$2.70\mltp10^{-4}$&$3.63\mltp10^{-4}$&$7.00\mltp10^{-3}$&$1.12\mltp10^{-2}$&$-$\\  
1000 & 0.229  &$7.96\mltp10^{-1}$

&$6.01\mltp10^{-4}$&$1.08\mltp10^{-4}$&$3.11\mltp10^{-4}$&$2.21\mltp10^{-3}$&$5.62\mltp10^{-3}$&$-$\\
1500 & 0.265 &$8.81\mltp10^{-1}$&$6.01\mltp10^{-4}$&$4.05\mltp10^{-5}$&$1.55\mltp10^{-4}$&$-$&$2.81\mltp10^{-3}$&$-$\\
2000 & 0.260 &$9.67\mltp10^{-1}$&$7.22\mltp10^{-4}$&$4.05\mltp10^{-5}$&$1.55\mltp10^{-4}$&$-$&$5.62\mltp10^{-3}$&$-$\\
2500 & 0.220 & $9.93\mltp10^{-1}$&$4.81\mltp10^{-4}$&$4.05\mltp10^{-5}$&$1.81\mltp10^{-4}$&$-$&$2.81\mltp10^{-3}$&$-$\\    
3000 &  0.181  &$9.97\mltp10^{-1}$&$3.61\mltp10^{-4}$&$4.05\mltp10^{-5}$&$1.30\mltp10^{-4}$&$-$&$2.81\mltp10^{-3}$&$-$\\
3500 &  0.155  &$9.99\mltp10^{-1}$&$4.01\mltp10^{-4}$&$5.39\mltp10^{-5}$&$1.04\mltp10^{-4}$&$-$&$2.81\mltp10^{-3}$&$-$\\
4000 &  0.118  &$9.99\mltp10^{-1}$&$2.81\mltp10^{-4}$&$4.58\mltp10^{-5}$&$7.77\mltp10^{-5}$&$-$&$2.81\mltp10^{-3}$&$-$\\	
\hline
\hline
\end{tabular}
}
\caption{
Selection efficiencies of BDT cuts for both signal and background processes at the HL-LHC with $\sqrt{s} =$ 14 TeV assuming different 
ALP masses, where the second column denotes the lower threshold of BDT responses, and ``$-$" means the number of events can be reduced to be negligible.
}
\label{tab:BDT}
\end{table*}

\acknowledgments
We thank Ye Lu and Yiheng Xiong for helpful discussions. Z.D. and K.W. are supported by the National Natural Science Foundation of China under grant no.~11905162, 
the Excellent Young Talents Program of the Wuhan University of Technology under grant no.~40122102, and the research program of the Wuhan University of Technology under grant no.~2020IB024.
Y.N.M. is supported by the National Natural Science Foundation of China under grant no.~12205227.
The simulation and analysis work of this article was completed with the computational cluster provided by the Theoretical Physics Group at the Department of Physics, School of Physics and Mechanics, Wuhan University of Technology.




\bibliography{Refs.bib}

\providecommand{\href}[2]{#2}\begingroup\raggedright\begin{thebibliography}{10}

\bibitem{Dine:2000cj}
M.~Dine, \emph{{TASI lectures on the strong CP problem}},  in
  \emph{{Theoretical Advanced Study Institute in Elementary Particle Physics
  (TASI 2000): Flavor Physics for the Millennium}}, pp.~349--369, 6, 2000
  [\href{https://arxiv.org/abs/hep-ph/0011376}{{\ttfamily hep-ph/0011376}}].

\bibitem{Kim:2008hd}
J.E.~Kim and G.~Carosi, \emph{{Axions and the Strong CP Problem}},
  \href{https://doi.org/10.1103/RevModPhys.82.557}{\emph{Rev. Mod. Phys.}
  {\bfseries 82} (2010) 557} [\href{https://arxiv.org/abs/0807.3125}{{\ttfamily
  0807.3125}}].

\bibitem{Kim:2009xp}
J.E.~Kim, \emph{{A Review on axions and the strong CP problem}},
  \href{https://doi.org/10.1063/1.3327743}{\emph{AIP Conf. Proc.} {\bfseries
  1200} (2010) 83} [\href{https://arxiv.org/abs/0909.3908}{{\ttfamily
  0909.3908}}].

\bibitem{Baker:2006ts}
C.A.~Baker et~al., \emph{{An Improved experimental limit on the electric dipole
  moment of the neutron}},
  \href{https://doi.org/10.1103/PhysRevLett.97.131801}{\emph{Phys. Rev. Lett.}
  {\bfseries 97} (2006) 131801}
  [\href{https://arxiv.org/abs/hep-ex/0602020}{{\ttfamily hep-ex/0602020}}].

\bibitem{Abel:2020pzs}
C.~Abel et~al., \emph{{Measurement of the Permanent Electric Dipole Moment of
  the Neutron}},
  \href{https://doi.org/10.1103/PhysRevLett.124.081803}{\emph{Phys. Rev. Lett.}
  {\bfseries 124} (2020) 081803}
  [\href{https://arxiv.org/abs/2001.11966}{{\ttfamily 2001.11966}}].

\bibitem{Peccei:1977hh}
R.D.~Peccei and H.R.~Quinn, \emph{{CP Conservation in the Presence of
  Instantons}}, \href{https://doi.org/10.1103/PhysRevLett.38.1440}{\emph{Phys.
  Rev. Lett.} {\bfseries 38} (1977) 1440}.

\bibitem{Peccei:1977ur}
R.D.~Peccei and H.R.~Quinn, \emph{{Constraints Imposed by CP Conservation in
  the Presence of Instantons}},
  \href{https://doi.org/10.1103/PhysRevD.16.1791}{\emph{Phys. Rev. D}
  {\bfseries 16} (1977) 1791}.

\bibitem{Galanti:2022ijh}
G.~Galanti and M.~Roncadelli, \emph{{Axion-like Particles Implications for
  High-Energy Astrophysics}},
  \href{https://doi.org/10.3390/universe8050253}{\emph{Universe} {\bfseries 8}
  (2022) 253} [\href{https://arxiv.org/abs/2205.00940}{{\ttfamily
  2205.00940}}].

\bibitem{Choi:2020rgn}
K.~Choi, S.H.~Im and C.~Sub~Shin, \emph{{Recent Progress in the Physics of
  Axions and Axion-Like Particles}},
  \href{https://doi.org/10.1146/annurev-nucl-120720-031147}{\emph{Ann. Rev.
  Nucl. Part. Sci.} {\bfseries 71} (2021) 225}
  [\href{https://arxiv.org/abs/2012.05029}{{\ttfamily 2012.05029}}].

\bibitem{Qiu:2024muo}
Q.~Qiu, Y.~Gao, H.-j.~Tian, K.~Wang, Z.~Wang and X.-M.~Yang, \emph{{Wide Binary
  Evaporation by Dark Solitons: Implications from the GAIA Catalog}},
  \href{https://arxiv.org/abs/2404.18099}{{\ttfamily 2404.18099}}.

\bibitem{Jaeckel:2015jla}
J.~Jaeckel and M.~Spannowsky, \emph{{Probing MeV to 90 GeV axion-like particles
  with LEP and LHC}},
  \href{https://doi.org/10.1016/j.physletb.2015.12.037}{\emph{Phys. Lett. B}
  {\bfseries 753} (2016) 482}
  [\href{https://arxiv.org/abs/1509.00476}{{\ttfamily 1509.00476}}].

\bibitem{Liu:2017zdh}
J.~Liu, L.-T.~Wang, X.-P.~Wang and W.~Xue, \emph{{Exposing the dark sector with
  future Z factories}},
  \href{https://doi.org/10.1103/PhysRevD.97.095044}{\emph{Phys. Rev. D}
  {\bfseries 97} (2018) 095044}
  [\href{https://arxiv.org/abs/1712.07237}{{\ttfamily 1712.07237}}].

\bibitem{Steinberg:2021iay}
N.~Steinberg and J.D.~Wells, \emph{{Axion-Like Particles at the ILC Giga-Z}},
  \href{https://doi.org/10.1007/JHEP08(2021)120}{\emph{JHEP} {\bfseries 08}
  (2021) 120} [\href{https://arxiv.org/abs/2101.00520}{{\ttfamily
  2101.00520}}].

\bibitem{Carra:2021ycg}
S.~Carra, V.~Goumarre, R.~Gupta, S.~Heim, B.~Heinemann, J.~Kuechler et~al.,
  \emph{{Constraining off-shell production of axionlike particles with
  Z\ensuremath{\gamma} and WW differential cross-section measurements}},
  \href{https://doi.org/10.1103/PhysRevD.104.092005}{\emph{Phys. Rev. D}
  {\bfseries 104} (2021) 092005}
  [\href{https://arxiv.org/abs/2106.10085}{{\ttfamily 2106.10085}}].

\bibitem{Bauer:2017ris}
M.~Bauer, M.~Neubert and A.~Thamm, \emph{{Collider Probes of Axion-Like
  Particles}}, \href{https://doi.org/10.1007/JHEP12(2017)044}{\emph{JHEP}
  {\bfseries 12} (2017) 044}
  [\href{https://arxiv.org/abs/1708.00443}{{\ttfamily 1708.00443}}].

\bibitem{Dolan:2017osp}
M.J.~Dolan, T.~Ferber, C.~Hearty, F.~Kahlhoefer and K.~Schmidt-Hoberg,
  \emph{{Revised constraints and Belle II sensitivity for visible and invisible
  axion-like particles}},
  \href{https://doi.org/10.1007/JHEP12(2017)094}{\emph{JHEP} {\bfseries 12}
  (2017) 094} [\href{https://arxiv.org/abs/1709.00009}{{\ttfamily
  1709.00009}}].

\bibitem{Bauer:2018uxu}
M.~Bauer, M.~Heiles, M.~Neubert and A.~Thamm, \emph{{Axion-Like Particles at
  Future Colliders}},
  \href{https://doi.org/10.1140/epjc/s10052-019-6587-9}{\emph{Eur. Phys. J. C}
  {\bfseries 79} (2019) 74} [\href{https://arxiv.org/abs/1808.10323}{{\ttfamily
  1808.10323}}].

\bibitem{Zhang:2021sio}
H.-Y.~Zhang, C.-X.~Yue, Y.-C.~Guo and S.~Yang, \emph{{Searching for axionlike
  particles at future electron-positron colliders}},
  \href{https://doi.org/10.1103/PhysRevD.104.096008}{\emph{Phys. Rev. D}
  {\bfseries 104} (2021) 096008}
  [\href{https://arxiv.org/abs/2103.05218}{{\ttfamily 2103.05218}}].

\bibitem{dEnterria:2021ljz}
D.~d'Enterria, \emph{{Collider constraints on axion-like particles}},  in
  \emph{{Workshop on Feebly Interacting Particles}}, 2, 2021
  [\href{https://arxiv.org/abs/2102.08971}{{\ttfamily 2102.08971}}].

\bibitem{Agrawal:2021dbo}
P.~Agrawal et~al., \emph{{Feebly-interacting particles: FIPs 2020 workshop
  report}}, \href{https://doi.org/10.1140/epjc/s10052-021-09703-7}{\emph{Eur.
  Phys. J. C} {\bfseries 81} (2021) 1015}
  [\href{https://arxiv.org/abs/2102.12143}{{\ttfamily 2102.12143}}].

\bibitem{Tian:2022rsi}
M.~Tian, Z.S.~Wang and K.~Wang, \emph{{Search for long-lived axions with far
  detectors at future lepton colliders}},
  \href{https://arxiv.org/abs/2201.08960}{{\ttfamily 2201.08960}}.

\bibitem{Ghebretinsaea:2022djg}
F.A.~Ghebretinsaea, Z.S.~Wang and K.~Wang, \emph{{Probing axion-like particles
  coupling to gluons at the LHC}},
  \href{https://doi.org/10.1007/JHEP07(2022)070}{\emph{JHEP} {\bfseries 07}
  (2022) 070} [\href{https://arxiv.org/abs/2203.01734}{{\ttfamily
  2203.01734}}].

\bibitem{Antel:2023hkf}
C.~Antel et~al., \emph{{Feebly-interacting particles: FIPs 2022 Workshop
  Report}}, \href{https://doi.org/10.1140/epjc/s10052-023-12168-5}{\emph{Eur.
  Phys. J. C} {\bfseries 83} (2023) 1122}
  [\href{https://arxiv.org/abs/2305.01715}{{\ttfamily 2305.01715}}].

\bibitem{Biswas:2023ksj}
T.~Biswas, \emph{{Probing the interactions of axion-like particles with
  electroweak bosons and the Higgs boson in the high energy regime at LHC}},
  \href{https://doi.org/10.1007/JHEP05(2024)081}{\emph{JHEP} {\bfseries 05}
  (2024) 081} [\href{https://arxiv.org/abs/2312.05992}{{\ttfamily
  2312.05992}}].

\bibitem{Lu:2024fxs}
Y.~Lu, Y.-n.~Mao, K.~Wang and Z.S.~Wang, \emph{{LAYCAST: LAYered CAvern Surface
  Tracker at future electron-positron colliders}},
  \href{https://arxiv.org/abs/2406.05770}{{\ttfamily 2406.05770}}.

\bibitem{BESIII:2022rzz}
{\scshape BESIII} collaboration, \emph{{Search for an axion-like particle in
  radiative J/\ensuremath{\psi} decays}},
  \href{https://doi.org/10.1016/j.physletb.2023.137698}{\emph{Phys. Lett. B}
  {\bfseries 838} (2023) 137698}
  [\href{https://arxiv.org/abs/2211.12699}{{\ttfamily 2211.12699}}].

\bibitem{Jiang:2023lnw}
{\scshape BESIII} collaboration, \emph{{ALPs searches at BESIII}},  in
  \emph{{57th Rencontres de Moriond on Electroweak Interactions and Unified
  Theories}}, 5, 2023 [\href{https://arxiv.org/abs/2305.08043}{{\ttfamily
  2305.08043}}].

\bibitem{Belle-II:2020jti}
{\scshape Belle-II} collaboration, \emph{{Search for Axion-Like Particles
  produced in $e^+e^-$ collisions at Belle II}},
  \href{https://doi.org/10.1103/PhysRevLett.125.161806}{\emph{Phys. Rev. Lett.}
  {\bfseries 125} (2020) 161806}
  [\href{https://arxiv.org/abs/2007.13071}{{\ttfamily 2007.13071}}].

\bibitem{ATLAS:2020hii}
{\scshape ATLAS} collaboration, \emph{{Measurement of light-by-light scattering
  and search for axion-like particles with 2.2 nb$^{-1}$ of Pb+Pb data with the
  ATLAS detector}}, \href{https://doi.org/10.1007/JHEP03(2021)243}{\emph{JHEP}
  {\bfseries 03} (2021) 243}
  [\href{https://arxiv.org/abs/2008.05355}{{\ttfamily 2008.05355}}].

\bibitem{CMS:2018erd}
{\scshape CMS} collaboration, \emph{{Evidence for light-by-light scattering and
  searches for axion-like particles in ultraperipheral PbPb collisions at
  $\sqrt{s_\mathrm{NN}} =$ 5.02 TeV}},
  \href{https://doi.org/10.1016/j.physletb.2019.134826}{\emph{Phys. Lett. B}
  {\bfseries 797} (2019) 134826}
  [\href{https://arxiv.org/abs/1810.04602}{{\ttfamily 1810.04602}}].

\bibitem{Dobrich:2015jyk}
B.~D\"obrich, J.~Jaeckel, F.~Kahlhoefer, A.~Ringwald and K.~Schmidt-Hoberg,
  \emph{{ALPtraum: ALP production in proton beam dump experiments}},
  \href{https://doi.org/10.1007/JHEP02(2016)018}{\emph{JHEP} {\bfseries 02}
  (2016) 018} [\href{https://arxiv.org/abs/1512.03069}{{\ttfamily
  1512.03069}}].

\bibitem{NA64:2020qwq}
{\scshape NA64} collaboration, \emph{{Search for Axionlike and Scalar Particles
  with the NA64 Experiment}},
  \href{https://doi.org/10.1103/PhysRevLett.125.081801}{\emph{Phys. Rev. Lett.}
  {\bfseries 125} (2020) 081801}
  [\href{https://arxiv.org/abs/2005.02710}{{\ttfamily 2005.02710}}].

\bibitem{ParticleDataGroup:2024cfk}
{\scshape Particle Data Group} collaboration, \emph{{Review of particle
  physics}}, \href{https://doi.org/10.1103/PhysRevD.110.030001}{\emph{Phys.
  Rev. D} {\bfseries 110} (2024) 030001}.

\bibitem{ALPlimits}
C.~O'HARE, \emph{{cajohare/AxionLimits: AxionLimits}},
  {\emph{https://doi.org/10.5281/zenodo.3932429} }.

\bibitem{Craig:2018kne}
N.~Craig, A.~Hook and S.~Kasko, \emph{{The Photophobic ALP}},
  \href{https://doi.org/10.1007/JHEP09(2018)028}{\emph{JHEP} {\bfseries 09}
  (2018) 028} [\href{https://arxiv.org/abs/1805.06538}{{\ttfamily
  1805.06538}}].

\bibitem{Fonseca:2018xzp}
N.~Fonseca, E.~Morgante and G.~Servant, \emph{{Higgs relaxation after
  inflation}}, \href{https://doi.org/10.1007/JHEP10(2018)020}{\emph{JHEP}
  {\bfseries 10} (2018) 020}
  [\href{https://arxiv.org/abs/1805.04543}{{\ttfamily 1805.04543}}].

\bibitem{Hook:2016mqo}
A.~Hook and G.~Marques-Tavares, \emph{{Relaxation from particle production}},
  \href{https://doi.org/10.1007/JHEP12(2016)101}{\emph{JHEP} {\bfseries 12}
  (2016) 101} [\href{https://arxiv.org/abs/1607.01786}{{\ttfamily
  1607.01786}}].

\bibitem{CDF:2022hxs}
{\scshape CDF} collaboration, \emph{{High-precision measurement of the $W$
  boson mass with the CDF II detector}},
  \href{https://doi.org/10.1126/science.abk1781}{\emph{Science} {\bfseries 376}
  (2022) 170}.

\bibitem{Aiko:2023trb}
M.~Aiko and M.~Endo, \emph{{Electroweak precision test of axion-like
  particles}}, \href{https://doi.org/10.1007/JHEP05(2023)147}{\emph{JHEP}
  {\bfseries 05} (2023) 147}
  [\href{https://arxiv.org/abs/2302.11377}{{\ttfamily 2302.11377}}].

\bibitem{Bonilla:2022pxu}
J.~Bonilla, I.~Brivio, J.~Machado-Rodr\'\i{}guez and J.F.~de~Troc\'oniz,
  \emph{{Nonresonant searches for axion-like particles in vector boson
  scattering processes at the LHC}},
  \href{https://doi.org/10.1007/JHEP06(2022)113}{\emph{JHEP} {\bfseries 06}
  (2022) 113} [\href{https://arxiv.org/abs/2202.03450}{{\ttfamily
  2202.03450}}].

\bibitem{Aiko:2024xiv}
M.~Aiko, M.~Endo and K.~Fridell, \emph{{Heavy photophobic ALP at the LHC}},
  \href{https://doi.org/10.1007/JHEP06(2024)194}{\emph{JHEP} {\bfseries 06}
  (2024) 194} [\href{https://arxiv.org/abs/2401.13323}{{\ttfamily
  2401.13323}}].

\bibitem{ATLAS:2016qjc}
{\scshape ATLAS} collaboration, \emph{{Measurements of $Z\gamma$ and
  $Z\gamma\gamma$ production in $pp$ collisions at $\sqrt{s}=$ 8 TeV with the
  ATLAS detector}},
  \href{https://doi.org/10.1103/PhysRevD.93.112002}{\emph{Phys. Rev. D}
  {\bfseries 93} (2016) 112002}
  [\href{https://arxiv.org/abs/1604.05232}{{\ttfamily 1604.05232}}].

\bibitem{ATLAS:2016jeu}
{\scshape ATLAS} collaboration, \emph{{Search for triboson $W^{\pm }W^{\pm
  }W^{\mp }$ production in $pp$ collisions at $\sqrt{s}=8$ $\text {TeV}$ with
  the ATLAS detector}},
  \href{https://doi.org/10.1140/epjc/s10052-017-4692-1}{\emph{Eur. Phys. J. C}
  {\bfseries 77} (2017) 141}
  [\href{https://arxiv.org/abs/1610.05088}{{\ttfamily 1610.05088}}].

\bibitem{CMS:2014cdf}
{\scshape CMS} collaboration, \emph{{Search for $WW \gamma$ and $WZ \gamma$
  production and constraints on anomalous quartic gauge couplings in $pp$
  collisions at $\sqrt s =$ 8 TeV}},
  \href{https://doi.org/10.1103/PhysRevD.90.032008}{\emph{Phys. Rev. D}
  {\bfseries 90} (2014) 032008}
  [\href{https://arxiv.org/abs/1404.4619}{{\ttfamily 1404.4619}}].

\bibitem{CMS:2020fqz}
{\scshape CMS} collaboration, \emph{{Evidence for electroweak production of
  four charged leptons and two jets in proton-proton collisions at $\sqrt {s}$
  = 13 TeV}}, \href{https://doi.org/10.1016/j.physletb.2020.135992}{\emph{Phys.
  Lett. B} {\bfseries 812} (2021) 135992}
  [\href{https://arxiv.org/abs/2008.07013}{{\ttfamily 2008.07013}}].

\bibitem{CMS:2021gme}
{\scshape CMS} collaboration, \emph{{Measurement of the electroweak production
  of Z$\gamma$ and two jets in proton-proton collisions at $\sqrt{s} =$ 13 TeV
  and constraints on anomalous quartic gauge couplings}},
  \href{https://doi.org/10.1103/PhysRevD.104.072001}{\emph{Phys. Rev. D}
  {\bfseries 104} (2021) 072001}
  [\href{https://arxiv.org/abs/2106.11082}{{\ttfamily 2106.11082}}].

\bibitem{CMS:2020ypo}
{\scshape CMS} collaboration, \emph{{Observation of electroweak production of
  W$\gamma$ with two jets in proton-proton collisions at $\sqrt {s}$ = 13
  TeV}}, \href{https://doi.org/10.1016/j.physletb.2020.135988}{\emph{Phys.
  Lett. B} {\bfseries 811} (2020) 135988}
  [\href{https://arxiv.org/abs/2008.10521}{{\ttfamily 2008.10521}}].

\bibitem{CMS:2020gfh}
{\scshape CMS} collaboration, \emph{{Measurements of production cross sections
  of WZ and same-sign WW boson pairs in association with two jets in
  proton-proton collisions at $\sqrt{s} =$ 13 TeV}},
  \href{https://doi.org/10.1016/j.physletb.2020.135710}{\emph{Phys. Lett. B}
  {\bfseries 809} (2020) 135710}
  [\href{https://arxiv.org/abs/2005.01173}{{\ttfamily 2005.01173}}].

\bibitem{CMS:2019mpq}
{\scshape CMS} collaboration, \emph{{Search for the production of
  W$^\pm$W$^\pm$W$^\mp$ events at $\sqrt{s} =$ 13 TeV}},
  \href{https://doi.org/10.1103/PhysRevD.100.012004}{\emph{Phys. Rev. D}
  {\bfseries 100} (2019) 012004}
  [\href{https://arxiv.org/abs/1905.04246}{{\ttfamily 1905.04246}}].

\bibitem{ATLAS:2021pdg}
{\scshape ATLAS} collaboration, \emph{{Observation of electroweak production of
  two jets in association with an isolated photon and missing transverse
  momentum, and search for a Higgs boson decaying into invisible particles at
  13~$\text {TeV}$ with the ATLAS detector}},
  \href{https://doi.org/10.1140/epjc/s10052-021-09878-z}{\emph{Eur. Phys. J. C}
  {\bfseries 82} (2022) 105}
  [\href{https://arxiv.org/abs/2109.00925}{{\ttfamily 2109.00925}}].

\bibitem{ATLAS:2023wqy}
{\scshape ATLAS} collaboration, \emph{{Search for the Z\ensuremath{\gamma}
  decay mode of new high-mass resonances in pp collisions at s=13 TeV with the
  ATLAS detector}},
  \href{https://doi.org/10.1016/j.physletb.2023.138394}{\emph{Phys. Lett. B}
  {\bfseries 848} (2024) 138394}
  [\href{https://arxiv.org/abs/2309.04364}{{\ttfamily 2309.04364}}].

\bibitem{Georgi:1986df}
H.~Georgi, D.B.~Kaplan and L.~Randall, \emph{{Manifesting the Invisible Axion
  at Low-energies}},
  \href{https://doi.org/10.1016/0370-2693(86)90688-X}{\emph{Phys. Lett. B}
  {\bfseries 169} (1986) 73}.

\bibitem{Brivio:2017ije}
I.~Brivio, M.B.~Gavela, L.~Merlo, K.~Mimasu, J.M.~No, R.~del Rey et~al.,
  \emph{{ALPs Effective Field Theory and Collider Signatures}},
  \href{https://doi.org/10.1140/epjc/s10052-017-5111-3}{\emph{Eur. Phys. J. C}
  {\bfseries 77} (2017) 572}
  [\href{https://arxiv.org/abs/1701.05379}{{\ttfamily 1701.05379}}].

\bibitem{Degrande:2011ua}
C.~Degrande, C.~Duhr, B.~Fuks, D.~Grellscheid, O.~Mattelaer and T.~Reiter,
  \emph{{UFO - The Universal FeynRules Output}},
  \href{https://doi.org/10.1016/j.cpc.2012.01.022}{\emph{Comput. Phys. Commun.}
  {\bfseries 183} (2012) 1201}
  [\href{https://arxiv.org/abs/1108.2040}{{\ttfamily 1108.2040}}].

\bibitem{Alwall:2014hca}
J.~Alwall, R.~Frederix, S.~Frixione, V.~Hirschi, F.~Maltoni, O.~Mattelaer
  et~al., \emph{{The automated computation of tree-level and next-to-leading
  order differential cross sections, and their matching to parton shower
  simulations}}, \href{https://doi.org/10.1007/JHEP07(2014)079}{\emph{JHEP}
  {\bfseries 07} (2014) 079} [\href{https://arxiv.org/abs/1405.0301}{{\ttfamily
  1405.0301}}].

\bibitem{Bierlich:2022pfr}
C.~Bierlich et~al., \emph{{A comprehensive guide to the physics and usage of
  PYTHIA 8.3}},
  \href{https://doi.org/10.21468/SciPostPhysCodeb.8}{\emph{SciPost Phys.
  Codeb.} {\bfseries 2022} (2022) 8}
  [\href{https://arxiv.org/abs/2203.11601}{{\ttfamily 2203.11601}}].

\bibitem{deFavereau:2013fsa}
{\scshape DELPHES 3} collaboration, \emph{{DELPHES 3, A modular framework for
  fast simulation of a generic collider experiment}},
  \href{https://doi.org/10.1007/JHEP02(2014)057}{\emph{JHEP} {\bfseries 02}
  (2014) 057} [\href{https://arxiv.org/abs/1307.6346}{{\ttfamily 1307.6346}}].

\bibitem{Cacciari:2011ma}
M.~Cacciari, G.P.~Salam and G.~Soyez, \emph{{FastJet User Manual}},
  \href{https://doi.org/10.1140/epjc/s10052-012-1896-2}{\emph{Eur. Phys. J. C}
  {\bfseries 72} (2012) 1896}
  [\href{https://arxiv.org/abs/1111.6097}{{\ttfamily 1111.6097}}].

\bibitem{Bauer:2020jbp}
M.~Bauer, M.~Neubert, S.~Renner, M.~Schnubel and A.~Thamm, \emph{{The
  Low-Energy Effective Theory of Axions and ALPs}},
  \href{https://doi.org/10.1007/JHEP04(2021)063}{\emph{JHEP} {\bfseries 04}
  (2021) 063} [\href{https://arxiv.org/abs/2012.12272}{{\ttfamily
  2012.12272}}].

\bibitem{Hocker:2007ht}
A.~Hocker et~al., \emph{{TMVA - Toolkit for Multivariate Data Analysis}},
  \href{https://arxiv.org/abs/physics/0703039}{{\ttfamily physics/0703039}}.

\bibitem{cowan2012discovery}
G.~Cowan, \emph{Discovery sensitivity for a counting experiment with background
  uncertainty}, .

\bibitem{ATLAS:2020yaz}
{\scshape ATLAS} collaboration, \emph{{Formulae for Estimating Significance}},
  .

\bibitem{Bhattiprolu:2020mwi}
P.N.~Bhattiprolu, S.P.~Martin and J.D.~Wells, \emph{{Criteria for projected
  discovery and exclusion sensitivities of counting experiments}},
  \href{https://doi.org/10.1140/epjc/s10052-020-08819-6}{\emph{Eur. Phys. J. C}
  {\bfseries 81} (2021) 123}
  [\href{https://arxiv.org/abs/2009.07249}{{\ttfamily 2009.07249}}].

\end{thebibliography}\endgroup
\bibliographystyle{JHEP.bst}

\end{document}